\documentclass[twocolumn,aps,prl,amssymb,amsmath,longbibliography,superscriptaddress]{revtex4-2}

\usepackage{hyperref}
\usepackage{graphicx}
\usepackage{amssymb}
\usepackage{bm}
\usepackage{xcolor}
\usepackage{ulem}

\newcommand{\F}{$\mathcal{F}$}  

\begin{document}

\title{Magnon-mediated terahertz spin transport in metallic Gd$\vert$Pt stacks}

\author{Oliver Gueckstock}
\email{Corresponding author: oliver.gueckstock@fu-berlin.de}
\affiliation{Fachbereich Physik, Freie Universit{\"a}t Berlin, Arnimallee 14,
	14195 Berlin, Germany}
\affiliation{Department of Physical Chemistry, Fritz Haber Institute of the Max Planck Society,  Faradayweg 4-6, 14195 Berlin, Germany}
\author{Tim Amrhein}
\affiliation{Fachbereich Physik, Freie Universit{\"a}t Berlin, Arnimallee 14,
14195 Berlin, Germany}
\author{Beatrice Andres}
\affiliation{Fachbereich Physik, Freie Universit{\"a}t Berlin, Arnimallee 14,
14195 Berlin, Germany}
\author{Pilar Jiménez-Cavero}
\affiliation{Centro Universitario de la Defensa, Academia General Militar, 50090 Zaragoza, Spain}
\author{Cornelius Gahl}
\affiliation{Fachbereich Physik, Freie Universit{\"a}t Berlin, Arnimallee 14,
	14195 Berlin, Germany}
\author{Tom S. Seifert}
\affiliation{Fachbereich Physik, Freie Universit{\"a}t Berlin, Arnimallee 14,
	14195 Berlin, Germany}
\author{Reza Rouzegar}
\affiliation{Fachbereich Physik, Freie Universit{\"a}t Berlin, Arnimallee 14,
	14195 Berlin, Germany}
\author{Ilie Radu}
\affiliation{Fachbereich Physik, Freie Universit{\"a}t Berlin, Arnimallee 14,
14195 Berlin, Germany}
\affiliation{European XFEL, Holzkoppel 4, 22869 Schenefeld, Germany}
\author{Irene Lucas}
\affiliation{Instituto de Nanociencia y Materiales de Aragón (INMA), Universidad de Zaragoza-CSIC, Mariano Esquillor, Edificio I+D, 50018 Zaragoza, Spain}
\affiliation{Departamento Física de la Materia Condensada, Universidad de Zaragoza, Pedro Cerbuna 12, 50009 Zaragoza, Spain}
\author{Marko Wietstruk}
\affiliation{Fachbereich Physik, Freie Universit{\"a}t Berlin, Arnimallee 14,
14195 Berlin, Germany}
\author{Luis Morellón}
\affiliation{Instituto de Nanociencia y Materiales de Aragón (INMA), Universidad de Zaragoza-CSIC, Mariano Esquillor, Edificio I+D, 50018 Zaragoza, Spain}
\affiliation{Departamento Física de la Materia Condensada, Universidad de Zaragoza, Pedro Cerbuna 12, 50009 Zaragoza, Spain}
\author{Martin Weinelt}
\affiliation{Fachbereich Physik, Freie Universit{\"a}t Berlin, Arnimallee 14,
14195 Berlin, Germany}
\author{Tobias Kampfrath}
\affiliation{Fachbereich Physik, Freie Universit{\"a}t Berlin, Arnimallee 14,
14195 Berlin, Germany}
\affiliation{Department of Physical Chemistry, Fritz Haber Institute of the Max Planck Society, Faradayweg 4-6, 14195 Berlin, Germany}
\author{Nele Thielemann-Kühn} 
\affiliation{Fachbereich Physik, Freie Universit{\"a}t Berlin, Arnimallee 14,
14195 Berlin, Germany}

\date{\today}

\begin{abstract}
We study femtosecond spin transport in a Gd$\vert$Pt stack induced by a laser pulse. Remarkably, the dynamics of the spin current from Gd to Pt suggests that its dominant driving force is the ultrafast spin Seebeck effect.
As the contribution of a transient spin voltage in the metal Gd is minor, Gd acts akin a magnetic insulator here. This view is supported by time- and spin-resolved photoemission, which indicates that a buildup of spin voltage is suppressed by exchange scattering, leading to similar amplitudes and relaxation rates of hot majority- and minority-spin electron populations. 

\end{abstract}

\pacs{Valid PACS appear here}

\maketitle


{\bf Introduction.} 
Two fundamental processes required for spintronic applications are spin transport and its detection \cite{Vedmedenko2020,Liang2021,Prinz1995,Zutic2004}. It is essential to push the speed of these operations to the ultrafast time scale to become competitive with terahertz (THz) clock rates achieved with other information carriers, such as photons \cite{Vedmedenko2020,Leitenstorfer2023}. 
In prototypical thin-film stacks \F$\vert$HM of a magnetic layer {\F} and a heavy-metal layer HM, THz spin transport can be triggered by excitation with a femtosecond (fs) laser pulse \cite{Malinowski2018}. The spin current that is injected into HM is typically detected by the inverse spin Hall effect, which is particularly large in Pt \cite{Sinova2015}.

Absorption of the pump pulse is known to induce two types of driving forces for spin transport from {\F} to HM: a difference in (A)~spin voltage or (B)~temperature between both layers. 
Typically, type~(A) is dominant in {\F}$\vert$HM stacks with metallic~{\F} containing $3d$ elements, where a spin-voltage imbalance induces spin transport carried by conduction electrons from {\F} to HM (pyrospintronic effect, PSE) \cite{Leitenstorfer2023,Wu2021,Feng2021,Bull2021,Rouzegar2022,Seifert2022,Liang2021,Schneider2022}. 
Type~(B) prevails for insulating {\F} with negligible pump absorptance because the {\F} electrons remain unexcited and cannot build up a spin voltage. However, the HM is excited, and the resulting temperature difference between the HM electrons and {\F} magnons drives a magnon-mediated spin current from {\F} to HM (interfacial ultrafast spin Seebeck effect, SSE) \cite{Seifert2018,Kholid2021,JimenezCavero2022}. 
In experiments on {\F}$\vert$Pt stacks with weakly conducting {\F}, the two driving forces (A) and (B) were successfully distinguished by their distinct dynamics \cite{JimenezCavero2022}. However, no signature of an ultrafast SSE has been observed for metallic~{\F}.

Remarkably, the ferromagnetic rare-earth 4$f$ metal Gd potentially offers both spin transport channels because it possesses localized $4f^7$ electronic states ($S = 7/2, L = 0$) at a binding energy of $E-E_{\rm{F}} \approx -8$\,eV and itinerant $(5d6s)^3$ valence electrons around the Fermi energy $E_{\rm{F}}$. They provide a magnetic moment of $7\mu_{\rm{B}}$ and $0.53\mu_{\rm{B}}$ per atom, respectively.
The $(5d6s)^3$ valence electrons are spin-polarized by exchange coupling to the $4f$ magnetic moments and, thus, mediate ferromagnetic order by the Ruderman-Kittel-Kasuya-Yosida (RKKY) interaction up to the bulk Curie temperature of $T_{\rm{C}}=297$\,K \cite{Graham1965,Farle1993}. 
Optical pump pulses with a typical photon energy of 1.5\,eV can directly excite the valence electrons but not the localized $4f$ spins \cite{Frietsch2020}. 

{\bf In this letter}, we study the origin of optically induced spin currents out of Gd thin films. 
We find that the spin-current dynamics in a Gd$\vert$Pt stack is nearly identical to that of well-known {\F}$\vert$Pt stacks with an insulating ferrimagnet~{\F}. 
We conclude that spin transport from Gd to Pt is mediated by magnon emission at the Gd$\vert$Pt interface, i.e., the THz SSE, even though Gd is highly metallic. Magnon generation is explained by the difference between the magnon temperature in Gd and the electron temperature in Pt and interfacial exchange interaction. We do not observe any signature of conduction-electron-mediated spin currents emanating Gd. In other words, Gd acts akin a magnetic insulator {\F} in {\F}$\vert$Pt stacks on ultrafast time scales.

This view is supported by complementary experiments using time-resolved photoelectron spectroscopy on Gd close to the Fermi edge. The amplitude and lifetime of hot-electron populations is found to be independent of their spin. This observation is attributed to rapid electron-electron exchange scattering, which leads to ultrafast alignment of excited majority- and minority-spin carrier densities during electron thermalization and, thus, a time-independent spin polarization after optical excitation of Gd. Our notion is supported by the slow decay of the Gd $5d$ exchange splitting (decay constant $\tau \approx 800$\,fs) upon optical excitation \cite{Carley2012,Frietsch2015}.

\begin{figure}
\includegraphics[width=1.0\columnwidth]{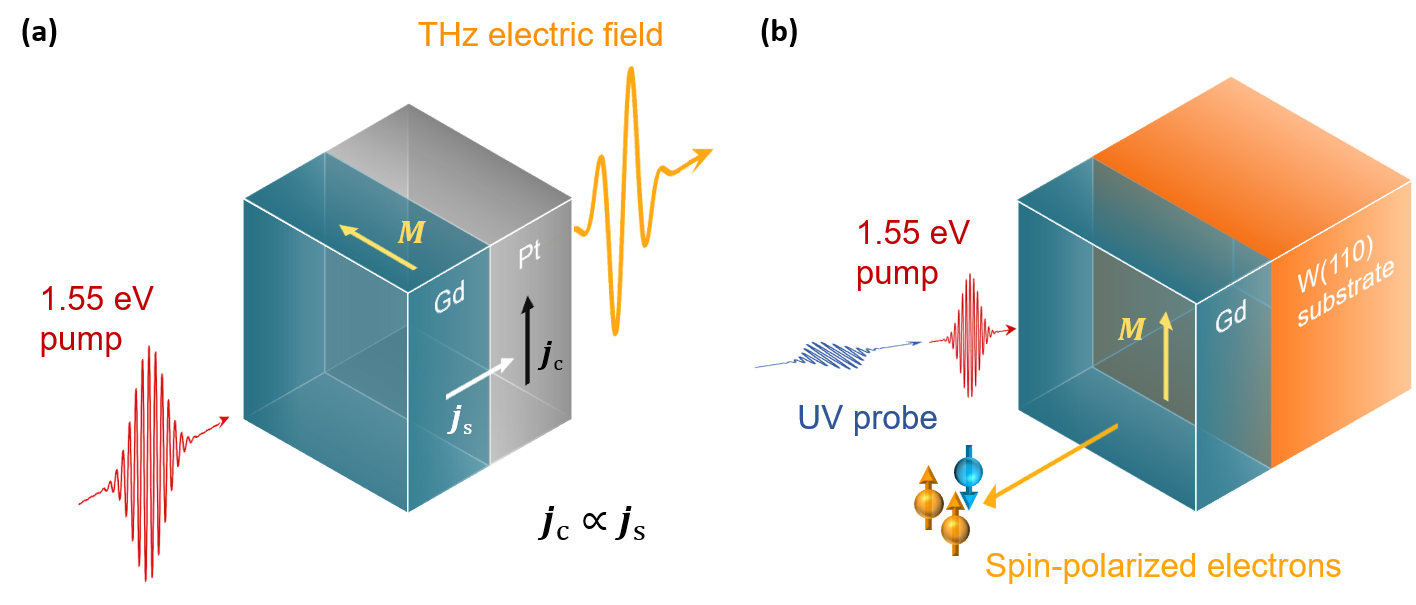}
\caption{Probing ultrafast spin dynamics in Gd induced by an optical femtosecond pump pulse of photon energy 1.55\,eV. (a)~Experiment~(1): Ultrafast spin transport in Gd(10\,nm)$\vert$Pt(2\,nm). The pump-induced spin current ${j_\text{s}}$ from Gd to Pt is in Pt converted into a transverse charge current ${j_\text{c}}$, which emits a THz electromagnetic pulse. (b)~Experiment~(2): Spin- and time-resolved photoemission of a Gd(10\,nm) film on W(110). The pump-induced hot-electron population dynamics in Gd is interrogated by a time-delayed ultraviolet (UV) probe pulse. Photo-emitted electrons are analyzed in terms of their energy $E - E_{\rm{F}}$ and spin polarization.}
\label{Figure_1}
\end{figure}

{\bf Experimental details.} 
We perform two experiments to address (1)~ultrafast spin transport out of Gd [Fig.~\ref{Figure_1}(a)] and (2)~ultrafast spin-resolved electron dynamics within Gd [Fig.~\ref{Figure_1}(b)] following excitation by a fs pump pulse. Experiment~(1) relies on THz-emission spectroscopy on a Y(5\,nm)$\vert$Gd(10\,nm)$\vert$Pt(2\,nm) stack on a 500\,$\mu$m-thick glass substrate grown in the DynaMaX sample preparation chamber at Helmholtz-Zentrum Berlin [Fig.~\ref{Figure_1}(a) and Supplemental Note~1.1]. An incident pump pulse from a Ti:sapphire laser oscillator (photon energy 1.55\,eV, nominal duration 10\,fs, repetition rate 80\,MHz, pulse energy 2\,nJ) triggers an out-of-plane spin current ${j}_\text{s}$ from Gd to Pt. The spin polarization is along the in-plane Gd magnetization $\boldsymbol{M}$. In Pt, the inverse spin Hall effect converts $\boldsymbol{j}_\text{s}$ into a time-dependent transverse charge current $\boldsymbol{j}_\text{c}$ that emits an electromagnetic pulse with frequencies extending into the THz range \cite{Seifert2022,Kampfrath2013,Seifert2016,Wu2021}. 

The THz electric field in the far-field is electro-optically sampled with a co-propagating probe pulse from the same laser \cite{Leitensdorfer1999,Kampfrath2007} in a ZnTe(110) electro-optic crystal (thickness 1\,mm). The temporal dynamics of the spin current $j_\text{s}$ is retrieved from the electro-optic signal $S(t)$ by numerical deconvolution of $S(t)=(H\ast E)(t)$ where $E(t)\propto j_\text{s}(t)$ denotes the THz electric field right behind the sample and $H(t)$ is the setup-specific transfer function that accounts for propagation and detector response \cite{Seifert2018,JimenezCavero2022}. 
An optical flow cryostat allows for sample temperatures down to $T_0 = 77$\,K. To saturate the sample magnetization $M$, we apply a homogeneous static external magnetic field with a strength of 95\,mT (Supplemental Note~2.2 and Fig.~S2).

For comparison to Gd$\vert$Pt, we also measure the reference samples Ref-PSE and Ref-SSE, which are model systems for the spin-transport driving forces (A) and (B), respectively. Ref-PSE is a commercially available spintronic trilayer W(2\,nm)$\vert$CoFeB(1.8\,nm)$\vert$Pt(2\,nm) (TeraSpinTec GmbH, Germany) \cite{Seifert2016}. Spin transport into the adjacent heavy metals W and Pt is driven by the spin voltage of CoFeB and carried by conduction electrons (PSE) \cite{Rouzegar2022,Fognini2017}. Ref-SSE is a $\gamma$-Fe$_2$O$_3$(10\,nm)$\vert$Pt(2.5\,nm) stack with insulating ferrimagnetic $\gamma$-Fe$_2$O$_3$ (maghemite) \cite{JimenezCavero2022} (Supplemental Note~1.3). Spin transport from maghemite to Pt is mediated by magnons and driven by the temperature difference of magnons in $\gamma$-Fe$_2$O$_3$ and of electrons in Pt (SSE) \cite{JimenezCavero2022,JimenezCavero2017, Seifert2018}.

In experiment~(2), we study spin-dependent hot-electron dynamics in a Gd(0001)(10\,nm) film on a W(110) single-crystal substrate by spin- and time-resolved photoelectron spectroscopy [Fig.~\ref{Figure_1}(b)]. Details on sample preparation and characterization can be found in Supplemental Notes~1 and~2. S-polarized 1.55-eV pump pulses from a Ti:Sapphire regenerative amplifier (Coherent RegA, 300 kHz) excite the sample, while frequency-quadrupled p-polarized probe pulses (6.2\,eV) induce photoelectron emission. About 10\% of the absorbed pump fluence of 3\,$\text{mJ}/\text{cm}^{2}$ are deposited in Gd (Supplemental Note~4.1). The pump-probe cross-correlation signal has a width $<120$\,fs. 

Photoelectrons are detected after a $90^{\circ}$ cylindrical sector analyzer (Focus CSA\,300). To resolve the spin polarization, we revert the magnetization of an Fe exchange-scattering detector and extract the signal component odd under this operation \cite{Winkelmann2008,Andres2015}. All spectra are taken under normal emission with an angular resolution of $\pm 2.5^{\circ}$ at a sample base temperature of 90\,K. The base pressure during measurement is $2 \cdot 10^{-11}$\,mbar. The remanent in-plane magnetization of Gd is reset by a magnetic-field pulse in between each laser shot using a free-standing coil along the Gd$[1100]$ direction. 

\begin{figure*}
\includegraphics[width=1.8\columnwidth]{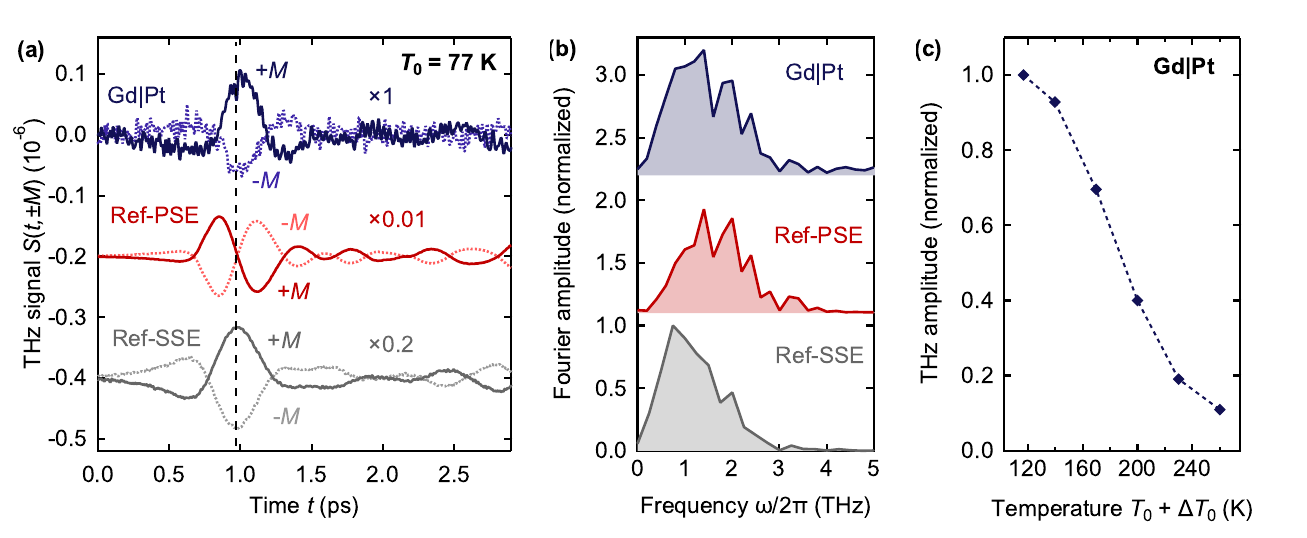}
\caption{THz-emission data and temperature dependence. (a)~Electro-optic signals $S(t,\pm M)$  from Gd$\vert$Pt for two opposite magnetization directions $\pm M$  (dark and light violet lines) at a set temperature $T_0$ of 77\,K. Reference THz-emission data are from the spin-voltage-driven THz emitter W$\vert$CoFeB$\vert$Pt (Ref-PSE; dark and light red) and from the temperature-difference-driven THz emitter $\gamma$-Fe$_2$O$_3$ $\vert$Pt (Ref-SSE; dark and light gray). Ref-PSE and Ref-SSE are measured at 77\,K and room temperature, respectively. Their signals are vertically rescaled and offset for clarity. (b)~Fourier-amplitude spectra of the odd-in-$M$ signals $S(t)$ of panel~(a). (c)~Normalized root-mean square (rms) amplitude of the waveforms $S(t)$ as function of the sample temperature $T_0+ \Delta T_0$. Here, $T_0$ is the set temperature, and $\Delta T_0 \sim 40$\,K is the estimated steady-state temperature increase due to the sample heating by the pump beam. Raw data are displayed in Supplemental Fig.~S3.}
\label{Figure_2}
\end{figure*}

{\bf Experiment~(1): Spin transport in Gd$\vert$Pt.}
Figure~\ref{Figure_2}(a) shows THz waveforms $S(t, \pm M)$ vs time $t$ for opposite magnetizations $\pm M$ from Gd$\vert$Pt (dark and light violet) and the references Ref-PSE (dark and light red) and Ref-SSE (dark and light gray). First, we note that, for all samples, the polarity of the waveform reverses when the magnetization $M$ is reversed. In the following, we focus on the odd-in-$M$ component $S(t)=[S(t,+M)-S(t,-M)]/2$, which is two orders of magnitude larger than the even-in-$M$ contribution. 

Second, the dynamics $S(t)$ of Gd$\vert$Pt and Ref-SSE is slower than that from Ref-PSE. This notion is supported by the Fourier-amplitude spectra of $S(t)$ [Fig.~\ref{Figure_2}(b)]. Interestingly, the waveforms from Gd$\vert$Pt and Ref-SSE have a very similar shape [Fig.~\ref{Figure_2}(a)].
Third, Fig.~\ref{Figure_2}(c) shows the THz amplitude from Gd$\vert$Pt as a function of the steady-state sample temperature $T_0+\Delta T_0$. The offset $\Delta T_0\sim 40$\,K accounts for continuous heating of the sample by the pump-pulse train (see Supplemental Note~3.2).

Figures~\ref{Figure_2}(c) and Supplemental Fig.~S4 reveal that the amplitude of $S(t)$ from Gd$\vert$Pt decreases monotonically with increasing temperature $T_0+\Delta T_0$. It becomes smaller than the noise level for $T_0+\Delta T_0 = 260$\,K. This value agrees roughly with the Curie temperature of Gd thin films, which was found to be about $T_\text{C}=280$\,K for a thickness of 10\,nm \cite{Farle1993}. Given the uncertainty of both $\Delta T_0$ and $T_\text{C}$ \cite{Farle1993}, the temperature dependence of the THz signal $S(t)$ [Fig.~\ref{Figure_2}(c)] is consistent with $S(t)$ originating from the ferromagnetic order of Gd. We further note the shape of the THz-emission amplitude vs $T_0$ may substantially deviate from known $M$-vs-$T_0$ behavior, as previously observed for Gd \cite{Melnikov04}, and for Y$_3$Fe$_5$O$_{12}\vert$Pt stacks in which exclusively magnon-mediated spin transport can occur \cite{Seifert2018} and 


\begin{figure*}
\includegraphics[width=1.5\columnwidth]{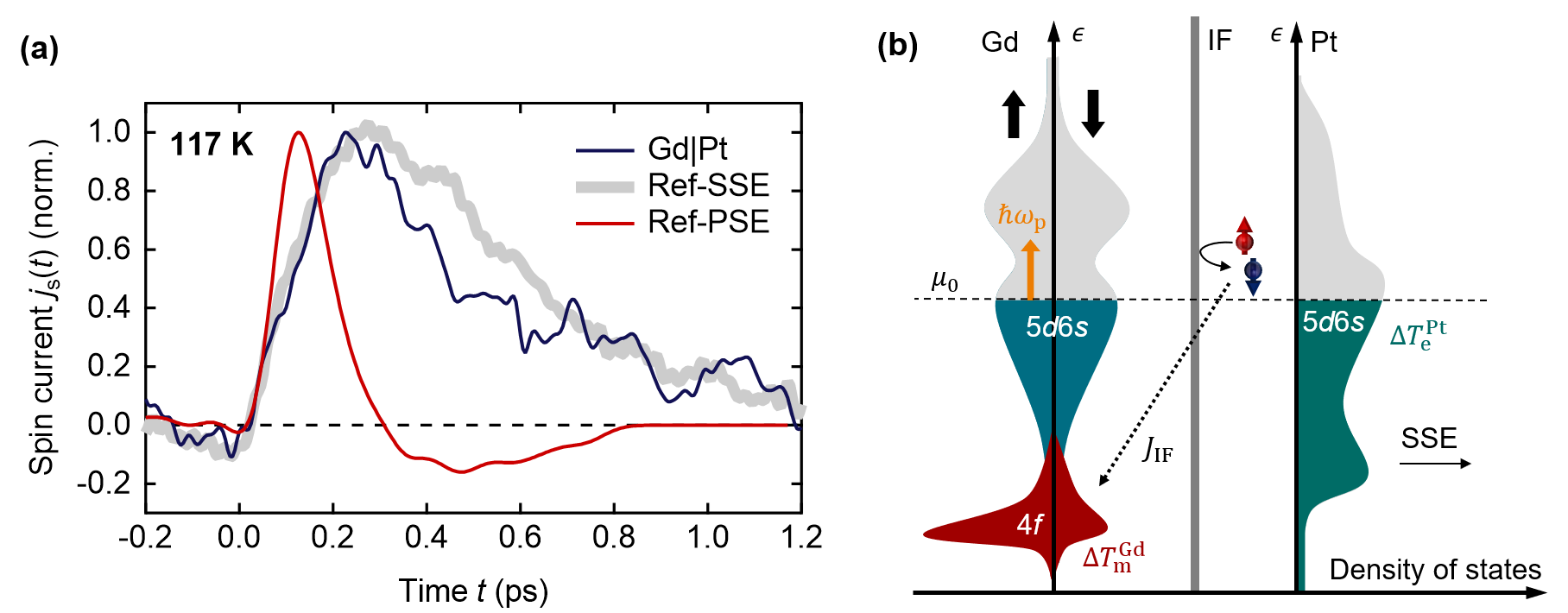}
\caption{Ultrafast spin-current dynamics and interpretation. (a)~Spin-current density $j_\text{s}(t)$ vs time~$t$ in Gd$\vert$Pt (dark blue) and the reference samples Ref-PSE (red) and Ref-SSE (gray). The Gd$\vert$Pt and Ref-PSE data are acquired at a sample temperature $T_0+\Delta T_0=120$\,K, whereas maghemite was measured at  $T_0=295\,$K. (b)~Proposed microscopic mechanism at the interface (IF) of Gd$\vert$Pt. The schematic of the spin-resolved electronic density of states indicates occupied $4f$ (red) and $5d6s$ states (blue) in Gd, and $5d6s$ states (green) in Pt. Unoccupied states above the Fermi energy $E_{\rm{F}}$ are shaded in gray. The orange arrow indicates the photon energy $\hbar\omega_{\text{p}} = 1.55$\,eV of the optical pump exciting Gd and Pt electrons. Spin transport from Gd to Pt is predominantly magnon-mediated, and its strength scales with the instantaneous temperature difference $T_{\text{e}}^{\text{Pt}}(t)-T_{\text{m}}^{\text{Gd}}(t)$ between hot electrons in Pt and cold magnons in Gd. Exchange coupling $J_{\text{IF}}$ between hot Pt electrons and Gd 4$f$ spins excites magnons in Gd and drives the spin current into the Pt layer (SSE).}
\label{Figure_3}
\end{figure*}

We use the THz signals of Fig.~\ref{Figure_2}(a) to extract the spin-current dynamics $j_\text{s}(t)$, which are compared to each other in Fig.~\ref{Figure_3}(a). Remarkably, the currents in Gd$\vert$Pt and Ref-SSE exhibit very similar dynamics and rise and decay substantially more slowly than in Ref-PSE, consistent with the raw signals in Fig.~\ref{Figure_2}(a,b). When the sample temperature is varied, we do not observe a significant change of $j_\text{s}(t)$ in Gd$\vert$Pt, apart from a global amplitude change (Supplemental Fig.~S4). Therefore, Fig.~\ref{Figure_3}(a) reveals that the spin-current dynamics  $j_\text{s}(t)$ for Gd$\vert$Pt very well agrees with that for maghemite$\vert$Pt, even though Gd is a metal and maghemite is an insulator. Such identical spin-current dynamics strongly indicates that the spin transport in Gd$\vert$Pt is predominantly driven by the magnonic ultrafast SSE.

To discuss the soundness of this notion, we consider the schematic of the electronic density of states of Gd and Pt in Fig.~\ref{Figure_3}(b). Ultrafast heating by the optical pump pulse induces a difference of (A)~the spin voltage and (B)~temperature between Gd and Pt layers. Both differences can give rise to a spin current (see introduction). 
(A)~A spin voltage $\mu_\text{s}$, arising from the elevated temperature of the $5d6s$ states close to the Fermi energy of Gd, and the resulting spin current $j_\text{s}\propto\mu_\text{s}$ would rise instantaneously once the pump pulse deposits energy in the electronic system \cite{Rouzegar2022}. Examples of such a rapid rise are the $j_\text{s}$ of Ref-PSE [Fig.~\ref{Figure_3}(a)] and {\F}$\vert$Pt stacks with {\F} made of itinerant $3d$-type ferromagnets 
\cite{Seifert2016,Wu2017,Sasaki2019,Schneider2022} or Heusler compounds
\cite{Bierhance2022,Sasaki2020,Yao2021,Heidtfeld2021,Gupta2021}. In strong contrast, the spin current in Gd$\vert$Pt rises more slowly and is, therefore, not predominantly driven by a spin voltage of the $5d6s$ electrons in Gd. 

(B)~Therefore, Fig.~\ref{Figure_3}(a) instead strongly suggests that $j_{\rm{s}}(t)$ in Gd$\vert$Pt arises from the temperature difference between Gd and Pt. This effect is well known in {\F}$\vert$Pt stacks with insulating ferrimagnetic {\F}, for instance, Y$_3$Fe$_5$O$_{12}$, Gd$_3$Fe$_5$O$_{12}$ and maghemite as in Ref-SSE \cite{Seifert2018,JimenezCavero2022,Kholid2021}. Here, the spin current $j_\text{s}(t)$ scales with the difference of the pump-induced changes in
the temperatures of electrons in Pt and magnons in Gd according to \cite{JimenezCavero2022}
\begin{equation}
j_{\text{s}}(t)\propto\Delta T_\text{e}^\text{Pt}(t) - \Delta T_\text{m}^\mathcal{F}(t). \label{Eq_1}
\end{equation}
Because the spin-current dynamics in Gd$\vert$Pt and Ref-SSE is identical, we conclude that the spin current in Gd$\vert$Pt is driven by the temperature difference $T_\text{e}^\text{Pt}(t)-T_\text{m}^\text{Gd}(t)$ of Pt electrons and Gd magnons. In the schematic of Fig.~\ref{Figure_3}(b), the magnons are carried by the $4f$ electrons at $E-E_\text{F}\approx-8$\,eV, which cannot be excited directly by the pump pulse (photon energy 1.55\,eV). Instead, exchange coupling between laser-heated Pt electrons and Gd 4$f$ spins excites magnons in Gd and drives spin transport into the Pt layer.

\begin{figure}
\includegraphics[width=1.0\columnwidth]{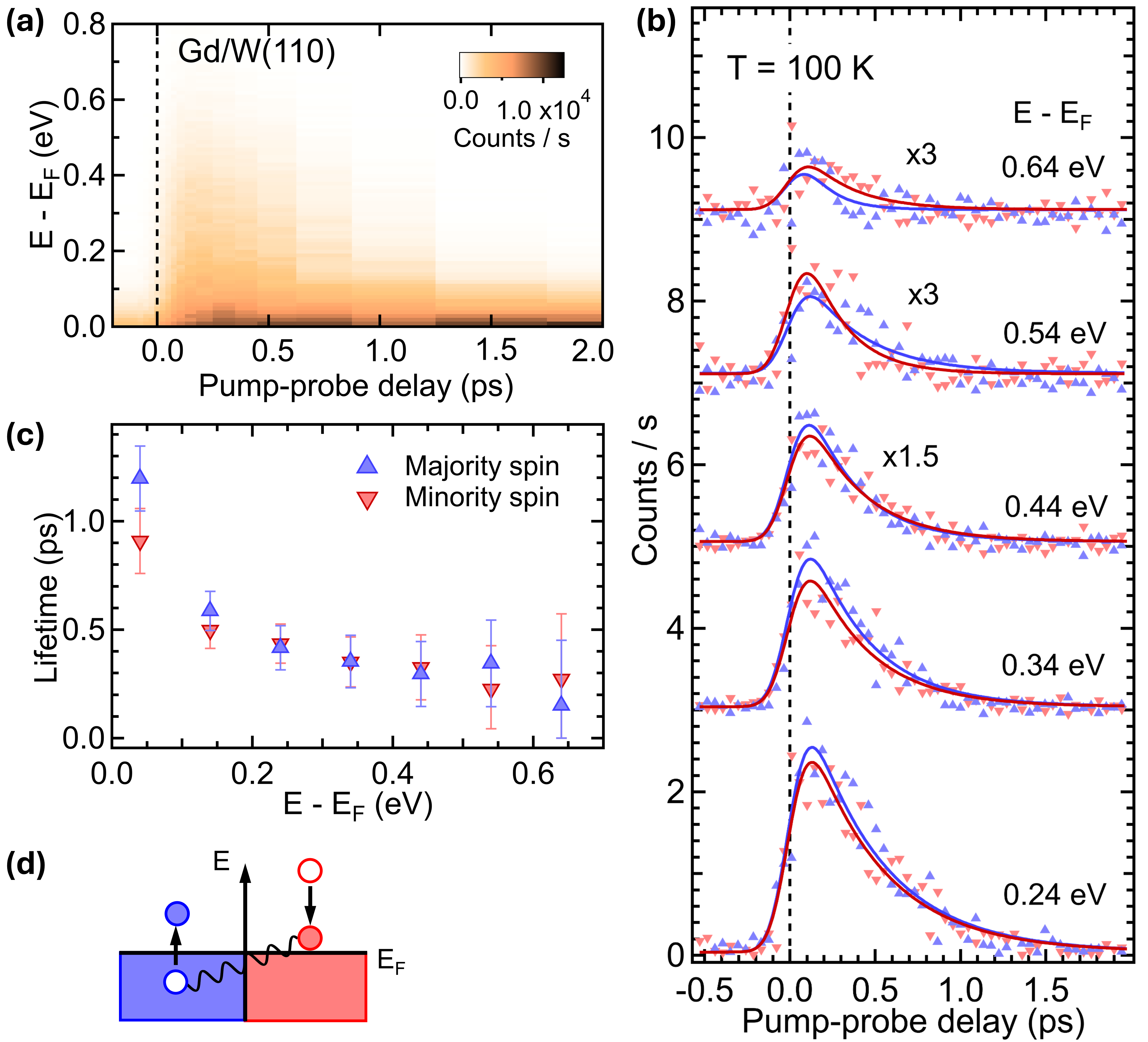}
\caption{Hot-electron dynamics in Gd. (a)~Photoemission spectra from epitaxial Gd(0001)(10\,nm) on W(110) at energies $E-E_{\rm{F}}>0$ as a function of pump-probe delay. (b)~Time-resolved traces for majority- and minority-spin electrons recorded at indicated energies $E-E_{\rm{F}}$. (c)~Decay times for both spin directions as determined from fitting the population decay. (d)~Schematic of inelastic electron-electron exchange scattering, which leads to an equilibration between majority- and minority-spin electron subsystems.}
\label{Figure_4}
\end{figure}

Our interpretation based on Eq.~(\ref{Eq_1}) and the almost identical spin currents $j_s (t)$ in Gd$\vert$Pt and maghemite$\vert$Pt has three implications. (i)~The dynamics of the transient electronic temperature change $\Delta T_{\text{e}}^{\text{Pt}}(t)$ in Pt is not substantially modified by the presence of metallic Gd. This implication is consistent with estimates of the electron-phonon relaxation in Gd$\vert$Pt based on the two-temperature model (Supplemental Note~3.4). (ii)~The temperature change $\Delta T_{\text{m}}^{\text{Gd}}$ of the Gd magnons is negligible compared to $\Delta T_{\text{e}}^{\rm{Pt}}$. This conclusion is consistent with the observation that $4f$ spins in optically excited Gd samples demagnetize on a time scale of about 15\,ps \cite{Frietsch2015}, much longer than the $\sim 1$\,ps interval of our experiment. (iii)~Eq.~(\ref{Eq_1}) neglects any spin voltage $\Delta \mu_{\text{s}}$ arising from optically excited Gd $5d6s$ electrons. 
Such a spin voltage could arise from either differences in density of states for spin up/down states at the Fermi edge or from population differences between up/down states. To further address implication (iii), we probe hot electrons in optically excited Gd(0001)(10\,nm) on W(110) by ultrafast spin- and time-resolved photoemission spectroscopy [Figs.~\ref{Figure_1}(b) and \ref{Figure_4}].

{\bf Experiment~(2): Electron dynamics in Gd$\vert$W}.
Figure\,\ref{Figure_4}(a) shows a photoelectron-intensity map for Gd recorded with a 6.2\,eV probe as a function of pump-probe delay. Corresponding time-resolved traces for majority and minority-spin components are depicted in Fig\,\ref{Figure_4}(b). Photoelectron spectroscopy is surface-sensitive. Because the escape depth of photoelectrons in Gd amounts to about 2.5\,\AA  \cite{Schoenhense1993}, we probe the spin-dependent electron dynamics at the Gd surface but not at the Gd/W interface. As evident from Figs.\,\ref{Figure_4}(b) and (c), we find comparable intensities and relaxation rates for majority- and minority-spin valence-electron populations excited above the Fermi level. 

The 1.55\,eV pump pulse generates a population of optically excited electrons, which equilibrates within 200\,fs \cite{Bovensiepen2007}.
The equilibration between the spin majority and minority channels occurs via inelastic electron-electron scattering, mediated by screened Coulomb interaction on a time scale of few femtoseconds \cite{Knorren2000,Zhukov2004,Goris2011,Kaltenborn2014}. It creates a hot population of thermalized secondary electrons. We note that, even at earlier times, the electron distribution is quite well described by a Fermi-Dirac distribution at temperature $T_{\rm{e}}$ \cite{Tengdin2018}. Figure\,\ref{Figure_4}(a) shows the decay of the hot Fermi tail as a function of energy $E - E_{\rm{F}}$ above the Fermi level. As the Coulomb interaction of two electrons with antiparallel spins is typically larger than that with parallel spins, scattering occurs more frequently between electrons of different spin. This so-called exchange scattering process \cite{Kirschner1985} is sketched in Fig.~\ref{Figure_4}(d). An electron decays in the minority spin channel and creates an electron-hole pair in the majority-spin channel or \textit{vice versa}. This process equalizes the phase space available for scattering in both spin channels \cite{Goris2011} and thereby aligns population differences {and relaxation rates} between hot majority- and minority-spin carriers on the ultrashort time scale of inelastic electron-electron scattering. 

{Unlike in the $3d$ metals, the majority- and minority-spin density of states (DOS) of the $5d$ valence electrons are comparable close to the Fermi level (see, \textit{e.g.}, Fig.\,S8 in \cite{Frietsch2020}). The valence electrons carry a magnetic moment of only $0.53\,\mu_{\rm{B}}$ per atom, and we observe little to no spin polarization at energies $E – E_{\rm{F}} > 0.2$\,eV in Fig.\,\ref{Figure_4}(b). The occupation of excited minority and majority spin electrons, $\Delta n^{\uparrow}$ and $\Delta n^{\downarrow}$, are comparable, and the generalized spin voltage $\propto \Delta n^{\uparrow} - \Delta n^{\downarrow}$ is negligible \cite{Rouzegar2022}. In addition, ultrafast electron-electron (exchange) scattering redistributes energy between majority and minority spin electrons at $E_{\rm{F}}$, resulting in an equilibration of their temperatures, $T_{\rm{e}\uparrow}=T_{\rm{e}\downarrow}$. Because the DOS at $E_{\rm{F}}$ is hardly spin dependent, the formation of a temperature-driven spin voltage is suppressed \cite{Rouzegar2022}.}

Slightly above the Fermi level (Supplemental Note~4.2, Figs.\,S6 and S7), the majority-spin valence-band photoemission is overlaid by contributions from the tail of the occupied majority-spin surface-state at a binding energy of $\approx 150$\,meV below $E_{\rm{F}}$ \cite{Bode1999,Andres2015}. In a recent work, we demonstrated optically induced spin transfer between the $d_{z^2}$ surface state and bulk states by a transient increase of the bulk exchange-splitting \cite{Bobowski2024}. In the present discussion, we can safely neglect the Gd surface state, since it will not form at the Gd$\vert$Pt interface and can be ruled out as source of the observed spin currents.

Our conclusion is further corroborated by the disparate dynamics of $5d$ and $4f$ magnetic momenta \cite{Frietsch2015}. While the exchange-splitting of the $5d$ valence bands exhibits a decay constant of $\approx 800$\,fs \cite{Frietsch2015,Bobowski2024}, $5d$ spin polarization and $4f$ magnetic moment are closely linked and change both with a slow 15\,ps time constant. The latter is attributed to rather weak spin-lattice coupling, based on the negligible orbital angular momentum ($L = 0$) of the half-filled Gd $4f^7$ shell \cite{Andres2015,Frietsch2020}. This point is also compatible with the assumption that the magnon temperature stays constant ($\Delta T_\text{m}^\text{Gd}=0$) over the 1\,ps time window of experiment~(1) [see Eq.\,(\ref{Eq_1})] and, thus, follows the case of Ref-SSE. If magnon generation in bulk Gd were ultrafast, the dynamics of $j_{\rm{s}}$ would be faster than in Ref-SSE. Furthermore, $5d$-$4f$ electron-electron scattering is negligible, since $4f$ multiplet excitations in Gd require higher photon energies than provided by the pump pulse \cite{Thielemann2024}. Therefore, compared to the build-up of the spin current at the Gd$\vert$Pt interface within 200\,fs, both the change in exchange splitting and magnon generation in pure Gd are slower. This notion is in contrast to $3d$-ferromagnets, where ultrafast demagnetization with typical time constants of $< 200$\,fs was reported \cite{Roth2012,Rhie2003,Stamm2007,Krauss2009,Kupier2014,Cinchetti2006,Weber2011,Carpene2008} {and spin current and demagnetization rate  are proportional, $j_{\rm{s}}(t) \propto \partial_t M(t)$ \cite{Rouzegar2022}. For Gd, this scenario would lead to a spin current $j_{\rm{s}}$ which increases much slower, i.e. with the 800\,fs decay-constant of the $5d$ exchange splitting. Instead we observe an increase of $j_{\rm{s}}$ within 200\,fs.}
All these peculiar properties of the Gd spin system support our observation that ultrafast spin currents in Gd are driven by magnons and, despite their low energies \cite{Jensen1991}, can only be excited at the Gd$\vert$Pt interface.

In conclusion, we observe that THz emission in a Gd$\vert$Pt bilayer is driven by the ultrafast interfacial spin Seebeck effect. Identical dynamics of the spin current in Gd$\vert$Pt and  $\gamma$-Fe$_2$O$_3\vert$Pt, with metallic Gd and insulating maghemite, are a hallmark of the absence of spin-polarized valence-band electrons around the Fermi level of Gd. We confirm this notion by spin- and time-resolved photoelectron spectroscopy showing comparable amplitudes and decay times of hot majority- and minority-spin electron populations. Magnons are generated at the Gd$\vert$Pt interface and drive an interfacial spin current, leading to demagnetization of Gd. Our observations corroborate the complex spin dynamics observed in Gd \cite{Carley2012,Andres2015,Frietsch2015,Frietsch2020} and, thus, suggest that all-optical switching dynamics in Gd/3d-metal alloys and layered systems \cite{Stanciu2007,Ostler2012,Kirilyuk2013} strongly depends on interfacial spin currents \cite{Graves2013,Beens2019,Iacocca2019}. Besides unravelling fundamental aspects of spin dynamics by THz-emission spectroscopy, combining Gd with $3d$ ferromagnets can open a novel approach to shape the emission spectrum and pulse duration in, \textit{e.g.}, Gd$\vert$Pt$\vert$3d-FM spintronic THz emitters.

\noindent
{\bf Acknowledgments.} {The authors appreciate the possibility to grow samples in the DynaMaX sample preparation chamber and are thankful for the PM3 beamtime granded by HZB to characterize the Gd$\vert$Pt bilayers.} The authors acknowledge funding by the German Research Foundation through the collaborative research center SFB TRR 227 “Ultrafast spin dynamics” (project ID 328545488, projects A01, A05 and B02) and the priority program SPP2314 “INTEREST” (project ID GE 3288 2-1, project ITISA), the Federal Ministry of Education and Research (BMBF) through Spinflash (project ID 05K22KE2), and the European Research Council through ERC-2023-AdG ORBITERA (grant No. 101142285).

\noindent
{\bf Conflict of Interest.} T.S.S. and T.K. are shareholders of TeraSpinTec GmbH, and T.S.S. is an employee of TeraSpinTec GmbH. The authors declare that they have no other competing interests.

\bibliographystyle{apsrev4-2}
\bibliography{GdPt}

\begin{thebibliography}{63}%
\makeatletter
\providecommand \@ifxundefined [1]{%
 \@ifx{#1\undefined}
}%
\providecommand \@ifnum [1]{%
 \ifnum #1\expandafter \@firstoftwo
 \else \expandafter \@secondoftwo
 \fi
}%
\providecommand \@ifx [1]{%
 \ifx #1\expandafter \@firstoftwo
 \else \expandafter \@secondoftwo
 \fi
}%
\providecommand \natexlab [1]{#1}%
\providecommand \enquote  [1]{``#1''}%
\providecommand \bibnamefont  [1]{#1}%
\providecommand \bibfnamefont [1]{#1}%
\providecommand \citenamefont [1]{#1}%
\providecommand \href@noop [0]{\@secondoftwo}%
\providecommand \href [0]{\begingroup \@sanitize@url \@href}%
\providecommand \@href[1]{\@@startlink{#1}\@@href}%
\providecommand \@@href[1]{\endgroup#1\@@endlink}%
\providecommand \@sanitize@url [0]{\catcode `\\12\catcode `\$12\catcode `\&12\catcode `\#12\catcode `\^12\catcode `\_12\catcode `\%12\relax}%
\providecommand \@@startlink[1]{}%
\providecommand \@@endlink[0]{}%
\providecommand \url  [0]{\begingroup\@sanitize@url \@url }%
\providecommand \@url [1]{\endgroup\@href {#1}{\urlprefix }}%
\providecommand \urlprefix  [0]{URL }%
\providecommand \Eprint [0]{\href }%
\providecommand \doibase [0]{https://doi.org/}%
\providecommand \selectlanguage [0]{\@gobble}%
\providecommand \bibinfo  [0]{\@secondoftwo}%
\providecommand \bibfield  [0]{\@secondoftwo}%
\providecommand \translation [1]{[#1]}%
\providecommand \BibitemOpen [0]{}%
\providecommand \bibitemStop [0]{}%
\providecommand \bibitemNoStop [0]{.\EOS\space}%
\providecommand \EOS [0]{\spacefactor3000\relax}%
\providecommand \BibitemShut  [1]{\csname bibitem#1\endcsname}%
\let\auto@bib@innerbib\@empty
\bibitem [{\citenamefont {Vedmedenko}\ \emph {et~al.}(2020)\citenamefont {Vedmedenko}, \citenamefont {Kawakami}, \citenamefont {Sheka}, \citenamefont {Gambardella}, \citenamefont {Kirilyuk}, \citenamefont {Hirohata}, \citenamefont {Binek}, \citenamefont {Chubykalo-Fesenko}, \citenamefont {Sanvito}, \citenamefont {Kirby}, \citenamefont {Grollier}, \citenamefont {Everschor-Sitte}, \citenamefont {Kampfrath}, \citenamefont {You},\ and\ \citenamefont {Berger}}]{Vedmedenko2020}%
  \BibitemOpen
  \bibfield  {author} {\bibinfo {author} {\bibfnamefont {E.~Y.}\ \bibnamefont {Vedmedenko}}, \bibinfo {author} {\bibfnamefont {R.~K.}\ \bibnamefont {Kawakami}}, \bibinfo {author} {\bibfnamefont {D.~D.}\ \bibnamefont {Sheka}}, \bibinfo {author} {\bibfnamefont {P.}~\bibnamefont {Gambardella}}, \bibinfo {author} {\bibfnamefont {A.}~\bibnamefont {Kirilyuk}}, \bibinfo {author} {\bibfnamefont {A.}~\bibnamefont {Hirohata}}, \bibinfo {author} {\bibfnamefont {C.}~\bibnamefont {Binek}}, \bibinfo {author} {\bibfnamefont {O.}~\bibnamefont {Chubykalo-Fesenko}}, \bibinfo {author} {\bibfnamefont {S.}~\bibnamefont {Sanvito}}, \bibinfo {author} {\bibfnamefont {B.~J.}\ \bibnamefont {Kirby}}, \bibinfo {author} {\bibfnamefont {J.}~\bibnamefont {Grollier}}, \bibinfo {author} {\bibfnamefont {K.}~\bibnamefont {Everschor-Sitte}}, \bibinfo {author} {\bibfnamefont {T.}~\bibnamefont {Kampfrath}}, \bibinfo {author} {\bibfnamefont {C.-Y.}\ \bibnamefont {You}},\ and\ \bibinfo {author} {\bibfnamefont {A.}~\bibnamefont {Berger}},\ }\href
  {https://doi.org/10.1088/1361-6463/ab9d98} {\bibfield  {journal} {\bibinfo  {journal} {Journal of Physics D: Applied Physics}\ }\textbf {\bibinfo {volume} {53}},\ \bibinfo {pages} {453001} (\bibinfo {year} {2020})}\BibitemShut {NoStop}%
\bibitem [{\citenamefont {Cheng}\ \emph {et~al.}(2021)\citenamefont {Cheng}, \citenamefont {Li}, \citenamefont {Zhao},\ and\ \citenamefont {Chia}}]{Liang2021}%
  \BibitemOpen
  \bibfield  {author} {\bibinfo {author} {\bibfnamefont {L.}~\bibnamefont {Cheng}}, \bibinfo {author} {\bibfnamefont {Z.}~\bibnamefont {Li}}, \bibinfo {author} {\bibfnamefont {D.}~\bibnamefont {Zhao}},\ and\ \bibinfo {author} {\bibfnamefont {E.~E.~M.}\ \bibnamefont {Chia}},\ }\href {https://doi.org/10.1063/5.0051217} {\bibfield  {journal} {\bibinfo  {journal} {APL Materials}\ }\textbf {\bibinfo {volume} {9}},\ \bibinfo {pages} {070902} (\bibinfo {year} {2021})}\BibitemShut {NoStop}%
\bibitem [{\citenamefont {Prinz}(1995)}]{Prinz1995}%
  \BibitemOpen
  \bibfield  {author} {\bibinfo {author} {\bibfnamefont {G.~A.}\ \bibnamefont {Prinz}},\ }\href {https://doi.org/10.1063/1.881459} {\bibfield  {journal} {\bibinfo  {journal} {Physics Today}\ }\textbf {\bibinfo {volume} {48}},\ \bibinfo {pages} {58} (\bibinfo {year} {1995})}\BibitemShut {NoStop}%
\bibitem [{\citenamefont {\ifmmode \check{Z}\else \v{Z}\fi{}uti\ifmmode~\acute{c}\else \'{c}\fi{}}\ \emph {et~al.}(2004)\citenamefont {\ifmmode \check{Z}\else \v{Z}\fi{}uti\ifmmode~\acute{c}\else \'{c}\fi{}}, \citenamefont {Fabian},\ and\ \citenamefont {Das~Sarma}}]{Zutic2004}%
  \BibitemOpen
  \bibfield  {author} {\bibinfo {author} {\bibfnamefont {I.}~\bibnamefont {\ifmmode \check{Z}\else \v{Z}\fi{}uti\ifmmode~\acute{c}\else \'{c}\fi{}}}, \bibinfo {author} {\bibfnamefont {J.}~\bibnamefont {Fabian}},\ and\ \bibinfo {author} {\bibfnamefont {S.}~\bibnamefont {Das~Sarma}},\ }\href {https://doi.org/10.1103/RevModPhys.76.323} {\bibfield  {journal} {\bibinfo  {journal} {Rev. Mod. Phys.}\ }\textbf {\bibinfo {volume} {76}},\ \bibinfo {pages} {323} (\bibinfo {year} {2004})}\BibitemShut {NoStop}%
\bibitem [{\citenamefont {Leitenstorfer}\ \emph {et~al.}(2023)\citenamefont {Leitenstorfer}, \citenamefont {Moskalenko}, \citenamefont {Kampfrath}, \citenamefont {Kono}, \citenamefont {Castro-Camus}, \citenamefont {Peng}, \citenamefont {Qureshi}, \citenamefont {Turchinovich}, \citenamefont {Tanaka}, \citenamefont {Markelz}, \citenamefont {Havenith}, \citenamefont {Hough}, \citenamefont {Joyce}, \citenamefont {Padilla}, \citenamefont {Zhou}, \citenamefont {Kim}, \citenamefont {Zhang}, \citenamefont {Jepsen}, \citenamefont {Dhillon}, \citenamefont {Vitiello}, \citenamefont {Linfield}, \citenamefont {Davies}, \citenamefont {Hoffmann}, \citenamefont {Lewis}, \citenamefont {Tonouchi}, \citenamefont {Klarskov}, \citenamefont {Seifert}, \citenamefont {Gerasimenko}, \citenamefont {Mihailovic}, \citenamefont {Huber}, \citenamefont {Boland}, \citenamefont {Mitrofanov}, \citenamefont {Dean}, \citenamefont {Ellison}, \citenamefont {Huggard}, \citenamefont {Rea}, \citenamefont {Walker}, \citenamefont {Leisawitz},
  \citenamefont {Gao}, \citenamefont {Li}, \citenamefont {Chen}, \citenamefont {Valušis}, \citenamefont {Wallace}, \citenamefont {Pickwell-MacPherson}, \citenamefont {Shang}, \citenamefont {Hesler}, \citenamefont {Ridler}, \citenamefont {Renaud}, \citenamefont {Kallfass}, \citenamefont {Nagatsuma}, \citenamefont {Zeitler}, \citenamefont {Arnone}, \citenamefont {Johnston},\ and\ \citenamefont {Cunningham}}]{Leitenstorfer2023}%
  \BibitemOpen
  \bibfield  {author} {\bibinfo {author} {\bibfnamefont {A.}~\bibnamefont {Leitenstorfer}}, \bibinfo {author} {\bibfnamefont {A.~S.}\ \bibnamefont {Moskalenko}}, \bibinfo {author} {\bibfnamefont {T.}~\bibnamefont {Kampfrath}}, \bibinfo {author} {\bibfnamefont {J.}~\bibnamefont {Kono}}, \bibinfo {author} {\bibfnamefont {E.}~\bibnamefont {Castro-Camus}}, \bibinfo {author} {\bibfnamefont {K.}~\bibnamefont {Peng}}, \bibinfo {author} {\bibfnamefont {N.}~\bibnamefont {Qureshi}}, \bibinfo {author} {\bibfnamefont {D.}~\bibnamefont {Turchinovich}}, \bibinfo {author} {\bibfnamefont {K.}~\bibnamefont {Tanaka}}, \bibinfo {author} {\bibfnamefont {A.~G.}\ \bibnamefont {Markelz}}, \bibinfo {author} {\bibfnamefont {M.}~\bibnamefont {Havenith}}, \bibinfo {author} {\bibfnamefont {C.}~\bibnamefont {Hough}}, \bibinfo {author} {\bibfnamefont {H.~J.}\ \bibnamefont {Joyce}}, \bibinfo {author} {\bibfnamefont {W.~J.}\ \bibnamefont {Padilla}}, \bibinfo {author} {\bibfnamefont {B.}~\bibnamefont {Zhou}}, \bibinfo {author} {\bibfnamefont
  {K.-Y.}\ \bibnamefont {Kim}}, \bibinfo {author} {\bibfnamefont {X.-C.}\ \bibnamefont {Zhang}}, \bibinfo {author} {\bibfnamefont {P.~U.}\ \bibnamefont {Jepsen}}, \bibinfo {author} {\bibfnamefont {S.}~\bibnamefont {Dhillon}}, \bibinfo {author} {\bibfnamefont {M.}~\bibnamefont {Vitiello}}, \bibinfo {author} {\bibfnamefont {E.}~\bibnamefont {Linfield}}, \bibinfo {author} {\bibfnamefont {A.~G.}\ \bibnamefont {Davies}}, \bibinfo {author} {\bibfnamefont {M.~C.}\ \bibnamefont {Hoffmann}}, \bibinfo {author} {\bibfnamefont {R.}~\bibnamefont {Lewis}}, \bibinfo {author} {\bibfnamefont {M.}~\bibnamefont {Tonouchi}}, \bibinfo {author} {\bibfnamefont {P.}~\bibnamefont {Klarskov}}, \bibinfo {author} {\bibfnamefont {T.~S.}\ \bibnamefont {Seifert}}, \bibinfo {author} {\bibfnamefont {Y.~A.}\ \bibnamefont {Gerasimenko}}, \bibinfo {author} {\bibfnamefont {D.}~\bibnamefont {Mihailovic}}, \bibinfo {author} {\bibfnamefont {R.}~\bibnamefont {Huber}}, \bibinfo {author} {\bibfnamefont {J.~L.}\ \bibnamefont {Boland}}, \bibinfo
  {author} {\bibfnamefont {O.}~\bibnamefont {Mitrofanov}}, \bibinfo {author} {\bibfnamefont {P.}~\bibnamefont {Dean}}, \bibinfo {author} {\bibfnamefont {B.~N.}\ \bibnamefont {Ellison}}, \bibinfo {author} {\bibfnamefont {P.~G.}\ \bibnamefont {Huggard}}, \bibinfo {author} {\bibfnamefont {S.~P.}\ \bibnamefont {Rea}}, \bibinfo {author} {\bibfnamefont {C.}~\bibnamefont {Walker}}, \bibinfo {author} {\bibfnamefont {D.~T.}\ \bibnamefont {Leisawitz}}, \bibinfo {author} {\bibfnamefont {J.~R.}\ \bibnamefont {Gao}}, \bibinfo {author} {\bibfnamefont {C.}~\bibnamefont {Li}}, \bibinfo {author} {\bibfnamefont {Q.}~\bibnamefont {Chen}}, \bibinfo {author} {\bibfnamefont {G.}~\bibnamefont {Valušis}}, \bibinfo {author} {\bibfnamefont {V.~P.}\ \bibnamefont {Wallace}}, \bibinfo {author} {\bibfnamefont {E.}~\bibnamefont {Pickwell-MacPherson}}, \bibinfo {author} {\bibfnamefont {X.}~\bibnamefont {Shang}}, \bibinfo {author} {\bibfnamefont {J.}~\bibnamefont {Hesler}}, \bibinfo {author} {\bibfnamefont {N.}~\bibnamefont {Ridler}},
  \bibinfo {author} {\bibfnamefont {C.~C.}\ \bibnamefont {Renaud}}, \bibinfo {author} {\bibfnamefont {I.}~\bibnamefont {Kallfass}}, \bibinfo {author} {\bibfnamefont {T.}~\bibnamefont {Nagatsuma}}, \bibinfo {author} {\bibfnamefont {J.~A.}\ \bibnamefont {Zeitler}}, \bibinfo {author} {\bibfnamefont {D.}~\bibnamefont {Arnone}}, \bibinfo {author} {\bibfnamefont {M.~B.}\ \bibnamefont {Johnston}},\ and\ \bibinfo {author} {\bibfnamefont {J.}~\bibnamefont {Cunningham}},\ }\href {https://doi.org/10.1088/1361-6463/acbe4c} {\bibfield  {journal} {\bibinfo  {journal} {Journal of Physics D: Applied Physics}\ }\textbf {\bibinfo {volume} {56}},\ \bibinfo {pages} {223001} (\bibinfo {year} {2023})}\BibitemShut {NoStop}%
\bibitem [{\citenamefont {Malinowski}\ \emph {et~al.}(2018)\citenamefont {Malinowski}, \citenamefont {Bergeard}, \citenamefont {Hehn},\ and\ \citenamefont {Mangin}}]{Malinowski2018}%
  \BibitemOpen
  \bibfield  {author} {\bibinfo {author} {\bibfnamefont {G.}~\bibnamefont {Malinowski}}, \bibinfo {author} {\bibfnamefont {N.}~\bibnamefont {Bergeard}}, \bibinfo {author} {\bibfnamefont {M.}~\bibnamefont {Hehn}},\ and\ \bibinfo {author} {\bibfnamefont {S.}~\bibnamefont {Mangin}},\ }\bibfield  {journal} {\bibinfo  {journal} {Eur. Phys. J. B}\ }\textbf {\bibinfo {volume} {91}},\ \href {https://doi.org/10.1140/epjb/e2018-80555-5} {10.1140/epjb/e2018-80555-5} (\bibinfo {year} {2018})\BibitemShut {NoStop}%
\bibitem [{\citenamefont {Sinova}\ \emph {et~al.}(2015)\citenamefont {Sinova}, \citenamefont {Valenzuela}, \citenamefont {Wunderlich}, \citenamefont {Back},\ and\ \citenamefont {Jungwirth}}]{Sinova2015}%
  \BibitemOpen
  \bibfield  {author} {\bibinfo {author} {\bibfnamefont {J.}~\bibnamefont {Sinova}}, \bibinfo {author} {\bibfnamefont {S.~O.}\ \bibnamefont {Valenzuela}}, \bibinfo {author} {\bibfnamefont {J.}~\bibnamefont {Wunderlich}}, \bibinfo {author} {\bibfnamefont {C.~H.}\ \bibnamefont {Back}},\ and\ \bibinfo {author} {\bibfnamefont {T.}~\bibnamefont {Jungwirth}},\ }\href {https://doi.org/10.1103/RevModPhys.87.1213} {\bibfield  {journal} {\bibinfo  {journal} {Rev. Mod. Phys.}\ }\textbf {\bibinfo {volume} {87}},\ \bibinfo {pages} {1213} (\bibinfo {year} {2015})}\BibitemShut {NoStop}%
\bibitem [{\citenamefont {Wu}\ \emph {et~al.}(2021)\citenamefont {Wu}, \citenamefont {Yaw~Ameyaw}, \citenamefont {Doty},\ and\ \citenamefont {Jungfleisch}}]{Wu2021}%
  \BibitemOpen
  \bibfield  {author} {\bibinfo {author} {\bibfnamefont {W.}~\bibnamefont {Wu}}, \bibinfo {author} {\bibfnamefont {C.}~\bibnamefont {Yaw~Ameyaw}}, \bibinfo {author} {\bibfnamefont {M.~F.}\ \bibnamefont {Doty}},\ and\ \bibinfo {author} {\bibfnamefont {M.~B.}\ \bibnamefont {Jungfleisch}},\ }\href {https://doi.org/10.1063/5.0057536} {\bibfield  {journal} {\bibinfo  {journal} {J. Appl. Phys.}\ }\textbf {\bibinfo {volume} {130}},\ \bibinfo {pages} {091101} (\bibinfo {year} {2021})}\BibitemShut {NoStop}%
\bibitem [{\citenamefont {Feng}\ \emph {et~al.}(2021)\citenamefont {Feng}, \citenamefont {Qiu}, \citenamefont {Wang}, \citenamefont {Zhang}, \citenamefont {Sun}, \citenamefont {Jin},\ and\ \citenamefont {Tan}}]{Feng2021}%
  \BibitemOpen
  \bibfield  {author} {\bibinfo {author} {\bibfnamefont {Z.}~\bibnamefont {Feng}}, \bibinfo {author} {\bibfnamefont {H.}~\bibnamefont {Qiu}}, \bibinfo {author} {\bibfnamefont {D.}~\bibnamefont {Wang}}, \bibinfo {author} {\bibfnamefont {C.}~\bibnamefont {Zhang}}, \bibinfo {author} {\bibfnamefont {S.}~\bibnamefont {Sun}}, \bibinfo {author} {\bibfnamefont {B.}~\bibnamefont {Jin}},\ and\ \bibinfo {author} {\bibfnamefont {W.}~\bibnamefont {Tan}},\ }\href {https://doi.org/10.1063/5.0037937} {\bibfield  {journal} {\bibinfo  {journal} {Journal of Applied Physics}\ }\textbf {\bibinfo {volume} {129}},\ \bibinfo {pages} {010901} (\bibinfo {year} {2021})}\BibitemShut {NoStop}%
\bibitem [{\citenamefont {Bull}\ \emph {et~al.}(2021)\citenamefont {Bull}, \citenamefont {Hewett}, \citenamefont {Ji}, \citenamefont {Lin}, \citenamefont {Thomson}, \citenamefont {Graham},\ and\ \citenamefont {Nutter}}]{Bull2021}%
  \BibitemOpen
  \bibfield  {author} {\bibinfo {author} {\bibfnamefont {C.}~\bibnamefont {Bull}}, \bibinfo {author} {\bibfnamefont {S.~M.}\ \bibnamefont {Hewett}}, \bibinfo {author} {\bibfnamefont {R.}~\bibnamefont {Ji}}, \bibinfo {author} {\bibfnamefont {C.-H.}\ \bibnamefont {Lin}}, \bibinfo {author} {\bibfnamefont {T.}~\bibnamefont {Thomson}}, \bibinfo {author} {\bibfnamefont {D.~M.}\ \bibnamefont {Graham}},\ and\ \bibinfo {author} {\bibfnamefont {P.~W.}\ \bibnamefont {Nutter}},\ }\href {https://doi.org/10.1063/5.0057511} {\bibfield  {journal} {\bibinfo  {journal} {APL Materials}\ }\textbf {\bibinfo {volume} {9}},\ \bibinfo {pages} {090701} (\bibinfo {year} {2021})}\BibitemShut {NoStop}%
\bibitem [{\citenamefont {Rouzegar}\ \emph {et~al.}(2022)\citenamefont {Rouzegar}, \citenamefont {Brandt}, \citenamefont {N\'advorn\'{\i}k}, \citenamefont {Reiss}, \citenamefont {Chekhov}, \citenamefont {Gueckstock}, \citenamefont {In}, \citenamefont {Wolf}, \citenamefont {Seifert}, \citenamefont {Brouwer}, \citenamefont {Woltersdorf},\ and\ \citenamefont {Kampfrath}}]{Rouzegar2022}%
  \BibitemOpen
  \bibfield  {author} {\bibinfo {author} {\bibfnamefont {R.}~\bibnamefont {Rouzegar}}, \bibinfo {author} {\bibfnamefont {L.}~\bibnamefont {Brandt}}, \bibinfo {author} {\bibfnamefont {L.~c.~v.}\ \bibnamefont {N\'advorn\'{\i}k}}, \bibinfo {author} {\bibfnamefont {D.~A.}\ \bibnamefont {Reiss}}, \bibinfo {author} {\bibfnamefont {A.~L.}\ \bibnamefont {Chekhov}}, \bibinfo {author} {\bibfnamefont {O.}~\bibnamefont {Gueckstock}}, \bibinfo {author} {\bibfnamefont {C.}~\bibnamefont {In}}, \bibinfo {author} {\bibfnamefont {M.}~\bibnamefont {Wolf}}, \bibinfo {author} {\bibfnamefont {T.~S.}\ \bibnamefont {Seifert}}, \bibinfo {author} {\bibfnamefont {P.~W.}\ \bibnamefont {Brouwer}}, \bibinfo {author} {\bibfnamefont {G.}~\bibnamefont {Woltersdorf}},\ and\ \bibinfo {author} {\bibfnamefont {T.}~\bibnamefont {Kampfrath}},\ }\href {https://doi.org/10.1103/PhysRevB.106.144427} {\bibfield  {journal} {\bibinfo  {journal} {Phys. Rev. B}\ }\textbf {\bibinfo {volume} {106}},\ \bibinfo {pages} {144427} (\bibinfo {year}
  {2022})}\BibitemShut {NoStop}%
\bibitem [{\citenamefont {Seifert}\ \emph {et~al.}(2022)\citenamefont {Seifert}, \citenamefont {Cheng}, \citenamefont {Wei}, \citenamefont {Kampfrath},\ and\ \citenamefont {Qi}}]{Seifert2022}%
  \BibitemOpen
  \bibfield  {author} {\bibinfo {author} {\bibfnamefont {T.~S.}\ \bibnamefont {Seifert}}, \bibinfo {author} {\bibfnamefont {L.}~\bibnamefont {Cheng}}, \bibinfo {author} {\bibfnamefont {Z.}~\bibnamefont {Wei}}, \bibinfo {author} {\bibfnamefont {T.}~\bibnamefont {Kampfrath}},\ and\ \bibinfo {author} {\bibfnamefont {J.}~\bibnamefont {Qi}},\ }\href {https://doi.org/10.1063/5.0080357} {\bibfield  {journal} {\bibinfo  {journal} {Appl. Phys. Lett.}\ }\textbf {\bibinfo {volume} {120}},\ \bibinfo {pages} {180401} (\bibinfo {year} {2022})}\BibitemShut {NoStop}%
\bibitem [{\citenamefont {Schneider}\ \emph {et~al.}(2022)\citenamefont {Schneider}, \citenamefont {Fix}, \citenamefont {Bensmann}, \citenamefont {Michaelis~de Vasconcellos}, \citenamefont {Albrecht},\ and\ \citenamefont {Bratschitsch}}]{Schneider2022}%
  \BibitemOpen
  \bibfield  {author} {\bibinfo {author} {\bibfnamefont {R.}~\bibnamefont {Schneider}}, \bibinfo {author} {\bibfnamefont {M.}~\bibnamefont {Fix}}, \bibinfo {author} {\bibfnamefont {J.}~\bibnamefont {Bensmann}}, \bibinfo {author} {\bibfnamefont {S.}~\bibnamefont {Michaelis~de Vasconcellos}}, \bibinfo {author} {\bibfnamefont {M.}~\bibnamefont {Albrecht}},\ and\ \bibinfo {author} {\bibfnamefont {R.}~\bibnamefont {Bratschitsch}},\ }\href {https://doi.org/10.1063/5.0076699} {\bibfield  {journal} {\bibinfo  {journal} {App. Phys. Lett.}\ }\textbf {\bibinfo {volume} {120}},\ \bibinfo {pages} {042404} (\bibinfo {year} {2022})}\BibitemShut {NoStop}%
\bibitem [{\citenamefont {Seifert}\ \emph {et~al.}(2018)\citenamefont {Seifert}, \citenamefont {Jaiswal}, \citenamefont {Barker}, \citenamefont {Weber}, \citenamefont {Razdolski}, \citenamefont {Cramer}, \citenamefont {Gueckstock}, \citenamefont {Maehrlein}, \citenamefont {Nadvornik}, \citenamefont {Watanabe}, \citenamefont {Ciccarelli}, \citenamefont {Melnikov}, \citenamefont {Jakob}, \citenamefont {Münzenberg}, \citenamefont {Goennenwein}, \citenamefont {Woltersdorf}, \citenamefont {Rethfeld}, \citenamefont {Brouwer}, \citenamefont {Wolf}, \citenamefont {Kl\"aui},\ and\ \citenamefont {Kampfrath}}]{Seifert2018}%
  \BibitemOpen
  \bibfield  {author} {\bibinfo {author} {\bibfnamefont {T.}~\bibnamefont {Seifert}}, \bibinfo {author} {\bibfnamefont {S.}~\bibnamefont {Jaiswal}}, \bibinfo {author} {\bibfnamefont {J.}~\bibnamefont {Barker}}, \bibinfo {author} {\bibfnamefont {S.}~\bibnamefont {Weber}}, \bibinfo {author} {\bibfnamefont {I.}~\bibnamefont {Razdolski}}, \bibinfo {author} {\bibfnamefont {J.}~\bibnamefont {Cramer}}, \bibinfo {author} {\bibfnamefont {O.}~\bibnamefont {Gueckstock}}, \bibinfo {author} {\bibfnamefont {S.~F.}\ \bibnamefont {Maehrlein}}, \bibinfo {author} {\bibfnamefont {L.}~\bibnamefont {Nadvornik}}, \bibinfo {author} {\bibfnamefont {S.}~\bibnamefont {Watanabe}}, \bibinfo {author} {\bibfnamefont {C.}~\bibnamefont {Ciccarelli}}, \bibinfo {author} {\bibfnamefont {A.}~\bibnamefont {Melnikov}}, \bibinfo {author} {\bibfnamefont {G.}~\bibnamefont {Jakob}}, \bibinfo {author} {\bibfnamefont {M.}~\bibnamefont {Münzenberg}}, \bibinfo {author} {\bibfnamefont {S.~T.~B.}\ \bibnamefont {Goennenwein}}, \bibinfo {author}
  {\bibfnamefont {G.}~\bibnamefont {Woltersdorf}}, \bibinfo {author} {\bibfnamefont {B.}~\bibnamefont {Rethfeld}}, \bibinfo {author} {\bibfnamefont {P.~W.}\ \bibnamefont {Brouwer}}, \bibinfo {author} {\bibfnamefont {M.}~\bibnamefont {Wolf}}, \bibinfo {author} {\bibfnamefont {M.}~\bibnamefont {Kl\"aui}},\ and\ \bibinfo {author} {\bibfnamefont {T.}~\bibnamefont {Kampfrath}},\ }\href {https://doi.org/10.1038/s41467-018-05135-2} {\bibfield  {journal} {\bibinfo  {journal} {Nat. Commun.}\ }\textbf {\bibinfo {volume} {9}},\ \bibinfo {pages} {2899} (\bibinfo {year} {2018})}\BibitemShut {NoStop}%
\bibitem [{\citenamefont {Kholid}\ \emph {et~al.}(2021)\citenamefont {Kholid}, \citenamefont {Hamara}, \citenamefont {Terschanski}, \citenamefont {Mertens}, \citenamefont {Bossini}, \citenamefont {Cinchetti}, \citenamefont {McKenzie-Sell}, \citenamefont {Patchett}, \citenamefont {Petit}, \citenamefont {Cowburn}, \citenamefont {Robinson}, \citenamefont {Barker},\ and\ \citenamefont {Ciccarelli}}]{Kholid2021}%
  \BibitemOpen
  \bibfield  {author} {\bibinfo {author} {\bibfnamefont {F.~N.}\ \bibnamefont {Kholid}}, \bibinfo {author} {\bibfnamefont {D.}~\bibnamefont {Hamara}}, \bibinfo {author} {\bibfnamefont {M.}~\bibnamefont {Terschanski}}, \bibinfo {author} {\bibfnamefont {F.}~\bibnamefont {Mertens}}, \bibinfo {author} {\bibfnamefont {D.}~\bibnamefont {Bossini}}, \bibinfo {author} {\bibfnamefont {M.}~\bibnamefont {Cinchetti}}, \bibinfo {author} {\bibfnamefont {L.}~\bibnamefont {McKenzie-Sell}}, \bibinfo {author} {\bibfnamefont {J.}~\bibnamefont {Patchett}}, \bibinfo {author} {\bibfnamefont {D.}~\bibnamefont {Petit}}, \bibinfo {author} {\bibfnamefont {R.}~\bibnamefont {Cowburn}}, \bibinfo {author} {\bibfnamefont {J.}~\bibnamefont {Robinson}}, \bibinfo {author} {\bibfnamefont {J.}~\bibnamefont {Barker}},\ and\ \bibinfo {author} {\bibfnamefont {C.}~\bibnamefont {Ciccarelli}},\ }\href {https://doi.org/10.1063/5.0050205} {\bibfield  {journal} {\bibinfo  {journal} {Appl. Phys. Lett.}\ }\textbf {\bibinfo {volume} {119}},\ \bibinfo
  {pages} {032401} (\bibinfo {year} {2021})}\BibitemShut {NoStop}%
\bibitem [{\citenamefont {Jim\'enez-Cavero}\ \emph {et~al.}(2022)\citenamefont {Jim\'enez-Cavero}, \citenamefont {Gueckstock}, \citenamefont {N\'advorn\'{\i}k}, \citenamefont {Lucas}, \citenamefont {Seifert}, \citenamefont {Wolf}, \citenamefont {Rouzegar}, \citenamefont {Brouwer}, \citenamefont {Becker}, \citenamefont {Jakob}, \citenamefont {Kl\"aui}, \citenamefont {Guo}, \citenamefont {Wan}, \citenamefont {Han}, \citenamefont {Jin}, \citenamefont {Zhao}, \citenamefont {Wu}, \citenamefont {Morell\'on},\ and\ \citenamefont {Kampfrath}}]{JimenezCavero2022}%
  \BibitemOpen
  \bibfield  {author} {\bibinfo {author} {\bibfnamefont {P.}~\bibnamefont {Jim\'enez-Cavero}}, \bibinfo {author} {\bibfnamefont {O.}~\bibnamefont {Gueckstock}}, \bibinfo {author} {\bibfnamefont {L.~c.~v.}\ \bibnamefont {N\'advorn\'{\i}k}}, \bibinfo {author} {\bibfnamefont {I.}~\bibnamefont {Lucas}}, \bibinfo {author} {\bibfnamefont {T.~S.}\ \bibnamefont {Seifert}}, \bibinfo {author} {\bibfnamefont {M.}~\bibnamefont {Wolf}}, \bibinfo {author} {\bibfnamefont {R.}~\bibnamefont {Rouzegar}}, \bibinfo {author} {\bibfnamefont {P.~W.}\ \bibnamefont {Brouwer}}, \bibinfo {author} {\bibfnamefont {S.}~\bibnamefont {Becker}}, \bibinfo {author} {\bibfnamefont {G.}~\bibnamefont {Jakob}}, \bibinfo {author} {\bibfnamefont {M.}~\bibnamefont {Kl\"aui}}, \bibinfo {author} {\bibfnamefont {C.}~\bibnamefont {Guo}}, \bibinfo {author} {\bibfnamefont {C.}~\bibnamefont {Wan}}, \bibinfo {author} {\bibfnamefont {X.}~\bibnamefont {Han}}, \bibinfo {author} {\bibfnamefont {Z.}~\bibnamefont {Jin}}, \bibinfo {author} {\bibfnamefont
  {H.}~\bibnamefont {Zhao}}, \bibinfo {author} {\bibfnamefont {D.}~\bibnamefont {Wu}}, \bibinfo {author} {\bibfnamefont {L.}~\bibnamefont {Morell\'on}},\ and\ \bibinfo {author} {\bibfnamefont {T.}~\bibnamefont {Kampfrath}},\ }\href {https://doi.org/10.1103/PhysRevB.105.184408} {\bibfield  {journal} {\bibinfo  {journal} {Phys. Rev. B}\ }\textbf {\bibinfo {volume} {105}},\ \bibinfo {pages} {184408} (\bibinfo {year} {2022})}\BibitemShut {NoStop}%
\bibitem [{\citenamefont {Graham}(1965)}]{Graham1965}%
  \BibitemOpen
  \bibfield  {author} {\bibinfo {author} {\bibfnamefont {J.}~\bibnamefont {Graham}, \bibfnamefont {C.~D.}},\ }\href {https://doi.org/10.1063/1.1714135} {\bibfield  {journal} {\bibinfo  {journal} {J. Appl. Phys.}\ }\textbf {\bibinfo {volume} {36}},\ \bibinfo {pages} {1135} (\bibinfo {year} {1965})}\BibitemShut {NoStop}%
\bibitem [{\citenamefont {Farle}\ \emph {et~al.}(1993)\citenamefont {Farle}, \citenamefont {Baberschke}, \citenamefont {Stetter}, \citenamefont {Aspelmeier},\ and\ \citenamefont {Gerhardter}}]{Farle1993}%
  \BibitemOpen
  \bibfield  {author} {\bibinfo {author} {\bibfnamefont {M.}~\bibnamefont {Farle}}, \bibinfo {author} {\bibfnamefont {K.}~\bibnamefont {Baberschke}}, \bibinfo {author} {\bibfnamefont {U.}~\bibnamefont {Stetter}}, \bibinfo {author} {\bibfnamefont {A.}~\bibnamefont {Aspelmeier}},\ and\ \bibinfo {author} {\bibfnamefont {F.}~\bibnamefont {Gerhardter}},\ }\href {https://doi.org/10.1103/PhysRevB.47.11571} {\bibfield  {journal} {\bibinfo  {journal} {Phys. Rev. B}\ }\textbf {\bibinfo {volume} {47}},\ \bibinfo {pages} {11571} (\bibinfo {year} {1993})}\BibitemShut {NoStop}%
\bibitem [{\citenamefont {Frietsch}\ \emph {et~al.}(2020)\citenamefont {Frietsch}, \citenamefont {Donges}, \citenamefont {Carley}, \citenamefont {Teichmann}, \citenamefont {Bowlan}, \citenamefont {D{\"o}brich}, \citenamefont {Carva}, \citenamefont {Legut}, \citenamefont {Oppeneer}, \citenamefont {Nowak},\ and\ \citenamefont {Weinelt}}]{Frietsch2020}%
  \BibitemOpen
  \bibfield  {author} {\bibinfo {author} {\bibfnamefont {B.}~\bibnamefont {Frietsch}}, \bibinfo {author} {\bibfnamefont {A.}~\bibnamefont {Donges}}, \bibinfo {author} {\bibfnamefont {R.}~\bibnamefont {Carley}}, \bibinfo {author} {\bibfnamefont {M.}~\bibnamefont {Teichmann}}, \bibinfo {author} {\bibfnamefont {J.}~\bibnamefont {Bowlan}}, \bibinfo {author} {\bibfnamefont {K.}~\bibnamefont {D{\"o}brich}}, \bibinfo {author} {\bibfnamefont {K.}~\bibnamefont {Carva}}, \bibinfo {author} {\bibfnamefont {D.}~\bibnamefont {Legut}}, \bibinfo {author} {\bibfnamefont {P.~M.}\ \bibnamefont {Oppeneer}}, \bibinfo {author} {\bibfnamefont {U.}~\bibnamefont {Nowak}},\ and\ \bibinfo {author} {\bibfnamefont {M.}~\bibnamefont {Weinelt}},\ }\href {https://doi.org/10.1126/sciadv.abb1601} {\bibfield  {journal} {\bibinfo  {journal} {Science Advances}\ }\textbf {\bibinfo {volume} {6}},\ \bibinfo {pages} {eabb1601} (\bibinfo {year} {2020})}\BibitemShut {NoStop}%
\bibitem [{\citenamefont {Carley}\ \emph {et~al.}(2012)\citenamefont {Carley}, \citenamefont {D{\"o}brich}, \citenamefont {Frietsch}, \citenamefont {Gahl}, \citenamefont {Teichmann}, \citenamefont {Schwarzkopf}, \citenamefont {Wernet},\ and\ \citenamefont {Weinelt}}]{Carley2012}%
  \BibitemOpen
  \bibfield  {author} {\bibinfo {author} {\bibfnamefont {R.}~\bibnamefont {Carley}}, \bibinfo {author} {\bibfnamefont {K.}~\bibnamefont {D{\"o}brich}}, \bibinfo {author} {\bibfnamefont {B.}~\bibnamefont {Frietsch}}, \bibinfo {author} {\bibfnamefont {C.}~\bibnamefont {Gahl}}, \bibinfo {author} {\bibfnamefont {M.}~\bibnamefont {Teichmann}}, \bibinfo {author} {\bibfnamefont {O.}~\bibnamefont {Schwarzkopf}}, \bibinfo {author} {\bibfnamefont {P.}~\bibnamefont {Wernet}},\ and\ \bibinfo {author} {\bibfnamefont {M.}~\bibnamefont {Weinelt}},\ }\href {https://doi.org/10.1103/PhysRevLett.109.057401} {\bibfield  {journal} {\bibinfo  {journal} {Phys. Rev. Lett.}\ }\textbf {\bibinfo {volume} {109}},\ \bibinfo {pages} {057401} (\bibinfo {year} {2012})}\BibitemShut {NoStop}%
\bibitem [{\citenamefont {Frietsch}\ \emph {et~al.}(2015)\citenamefont {Frietsch}, \citenamefont {Bowlan}, \citenamefont {Carley}, \citenamefont {Teichmann}, \citenamefont {Wienholdt}, \citenamefont {Hinzke}, \citenamefont {Nowak}, \citenamefont {Carva}, \citenamefont {Oppeneer},\ and\ \citenamefont {Weinelt}}]{Frietsch2015}%
  \BibitemOpen
  \bibfield  {author} {\bibinfo {author} {\bibfnamefont {B.}~\bibnamefont {Frietsch}}, \bibinfo {author} {\bibfnamefont {J.}~\bibnamefont {Bowlan}}, \bibinfo {author} {\bibfnamefont {R.}~\bibnamefont {Carley}}, \bibinfo {author} {\bibfnamefont {M.}~\bibnamefont {Teichmann}}, \bibinfo {author} {\bibfnamefont {S.}~\bibnamefont {Wienholdt}}, \bibinfo {author} {\bibfnamefont {D.}~\bibnamefont {Hinzke}}, \bibinfo {author} {\bibfnamefont {U.}~\bibnamefont {Nowak}}, \bibinfo {author} {\bibfnamefont {K.}~\bibnamefont {Carva}}, \bibinfo {author} {\bibfnamefont {P.~M.}\ \bibnamefont {Oppeneer}},\ and\ \bibinfo {author} {\bibfnamefont {M.}~\bibnamefont {Weinelt}},\ }\href {https://doi.org/10.1038/ncomms9262} {\bibfield  {journal} {\bibinfo  {journal} {Nat. Commun.}\ }\textbf {\bibinfo {volume} {6}},\ \bibinfo {pages} {8262} (\bibinfo {year} {2015})}\BibitemShut {NoStop}%
\bibitem [{\citenamefont {Kampfrath}\ \emph {et~al.}(2013)\citenamefont {Kampfrath}, \citenamefont {Battiato}, \citenamefont {Maldonado}, \citenamefont {P.}, \citenamefont {Eilers}, \citenamefont {N{\"o}tzold}, \citenamefont {Mährlein}, \citenamefont {Zbarsky},\ and\ \citenamefont {Freimuth}}]{Kampfrath2013}%
  \BibitemOpen
  \bibfield  {author} {\bibinfo {author} {\bibfnamefont {T.}~\bibnamefont {Kampfrath}}, \bibinfo {author} {\bibfnamefont {M.}~\bibnamefont {Battiato}}, \bibinfo {author} {\bibfnamefont {P.}~\bibnamefont {Maldonado}}, \bibinfo {author} {\bibfnamefont {M.}~\bibnamefont {P.}}, \bibinfo {author} {\bibfnamefont {G.}~\bibnamefont {Eilers}}, \bibinfo {author} {\bibfnamefont {J.}~\bibnamefont {N{\"o}tzold}}, \bibinfo {author} {\bibfnamefont {S.}~\bibnamefont {Mährlein}}, \bibinfo {author} {\bibfnamefont {V.}~\bibnamefont {Zbarsky}},\ and\ \bibinfo {author} {\bibfnamefont {F.}~\bibnamefont {Freimuth}},\ }\href {https://doi.org/10.1038/nnano.2013.43} {\bibfield  {journal} {\bibinfo  {journal} {Nat. Nanotechnol.}\ }\textbf {\bibinfo {volume} {8}},\ \bibinfo {pages} {256} (\bibinfo {year} {2013})}\BibitemShut {NoStop}%
\bibitem [{\citenamefont {Seifert}\ \emph {et~al.}(2016)\citenamefont {Seifert}, \citenamefont {Jaiswal}, \citenamefont {Martens}, \citenamefont {Hannegan}, \citenamefont {Braun}, \citenamefont {Maldonado}, \citenamefont {Freimuth}, \citenamefont {Kronenberg}, \citenamefont {Henrizi}, \citenamefont {Radu}, \citenamefont {Beaurepaire}, \citenamefont {Mokrousov}, \citenamefont {Oppeneer}, \citenamefont {Jourdan}, \citenamefont {Jakob}, \citenamefont {Turchinovich}, \citenamefont {Hayden}, \citenamefont {Wolf}, \citenamefont {Münzenberg}, \citenamefont {Kläui},\ and\ \citenamefont {Kampfrath}}]{Seifert2016}%
  \BibitemOpen
  \bibfield  {author} {\bibinfo {author} {\bibfnamefont {T.}~\bibnamefont {Seifert}}, \bibinfo {author} {\bibfnamefont {S.}~\bibnamefont {Jaiswal}}, \bibinfo {author} {\bibfnamefont {U.}~\bibnamefont {Martens}}, \bibinfo {author} {\bibfnamefont {J.}~\bibnamefont {Hannegan}}, \bibinfo {author} {\bibfnamefont {L.}~\bibnamefont {Braun}}, \bibinfo {author} {\bibfnamefont {P.}~\bibnamefont {Maldonado}}, \bibinfo {author} {\bibfnamefont {F.}~\bibnamefont {Freimuth}}, \bibinfo {author} {\bibfnamefont {A.}~\bibnamefont {Kronenberg}}, \bibinfo {author} {\bibfnamefont {J.}~\bibnamefont {Henrizi}}, \bibinfo {author} {\bibfnamefont {I.}~\bibnamefont {Radu}}, \bibinfo {author} {\bibfnamefont {E.}~\bibnamefont {Beaurepaire}}, \bibinfo {author} {\bibfnamefont {Y.}~\bibnamefont {Mokrousov}}, \bibinfo {author} {\bibfnamefont {P.~M.}\ \bibnamefont {Oppeneer}}, \bibinfo {author} {\bibfnamefont {M.}~\bibnamefont {Jourdan}}, \bibinfo {author} {\bibfnamefont {G.}~\bibnamefont {Jakob}}, \bibinfo {author} {\bibfnamefont
  {D.}~\bibnamefont {Turchinovich}}, \bibinfo {author} {\bibfnamefont {L.~M.}\ \bibnamefont {Hayden}}, \bibinfo {author} {\bibfnamefont {M.}~\bibnamefont {Wolf}}, \bibinfo {author} {\bibfnamefont {M.}~\bibnamefont {Münzenberg}}, \bibinfo {author} {\bibfnamefont {M.}~\bibnamefont {Kläui}},\ and\ \bibinfo {author} {\bibfnamefont {T.}~\bibnamefont {Kampfrath}},\ }\href {https://doi.org/10.1038/nphoton.2016.91} {\bibfield  {journal} {\bibinfo  {journal} {Nat. Photonics}\ }\textbf {\bibinfo {volume} {10}},\ \bibinfo {pages} {483} (\bibinfo {year} {2016})}\BibitemShut {NoStop}%
\bibitem [{\citenamefont {Leitenstorfer}\ \emph {et~al.}(1999)\citenamefont {Leitenstorfer}, \citenamefont {Hunsche}, \citenamefont {Shah}, \citenamefont {Nuss},\ and\ \citenamefont {Knox}}]{Leitensdorfer1999}%
  \BibitemOpen
  \bibfield  {author} {\bibinfo {author} {\bibfnamefont {A.}~\bibnamefont {Leitenstorfer}}, \bibinfo {author} {\bibfnamefont {S.}~\bibnamefont {Hunsche}}, \bibinfo {author} {\bibfnamefont {J.}~\bibnamefont {Shah}}, \bibinfo {author} {\bibfnamefont {M.~C.}\ \bibnamefont {Nuss}},\ and\ \bibinfo {author} {\bibfnamefont {W.~H.}\ \bibnamefont {Knox}},\ }\href {https://doi.org/10.1063/1.123601} {\bibfield  {journal} {\bibinfo  {journal} {Appl. Phys. Lett.}\ }\textbf {\bibinfo {volume} {74}},\ \bibinfo {pages} {1516} (\bibinfo {year} {1999})}\BibitemShut {NoStop}%
\bibitem [{\citenamefont {Stumpf}\ \emph {et~al.}(2007)\citenamefont {Stumpf}, \citenamefont {Fujita}, \citenamefont {Yamaguchi}, \citenamefont {Asano},\ and\ \citenamefont {Noda}}]{Kampfrath2007}%
  \BibitemOpen
  \bibfield  {author} {\bibinfo {author} {\bibfnamefont {W.~C.}\ \bibnamefont {Stumpf}}, \bibinfo {author} {\bibfnamefont {M.}~\bibnamefont {Fujita}}, \bibinfo {author} {\bibfnamefont {M.}~\bibnamefont {Yamaguchi}}, \bibinfo {author} {\bibfnamefont {T.}~\bibnamefont {Asano}},\ and\ \bibinfo {author} {\bibfnamefont {S.}~\bibnamefont {Noda}},\ }\href {https://doi.org/10.1063/1.2746059} {\bibfield  {journal} {\bibinfo  {journal} {Appl. Phys. Lett.}\ }\textbf {\bibinfo {volume} {90}},\ \bibinfo {pages} {231101} (\bibinfo {year} {2007})}\BibitemShut {NoStop}%
\bibitem [{\citenamefont {Fognini}\ \emph {et~al.}(2017)\citenamefont {Fognini}, \citenamefont {Michlmayr}, \citenamefont {Vaterlaus},\ and\ \citenamefont {Acremann}}]{Fognini2017}%
  \BibitemOpen
  \bibfield  {author} {\bibinfo {author} {\bibfnamefont {A.}~\bibnamefont {Fognini}}, \bibinfo {author} {\bibfnamefont {T.~U.}\ \bibnamefont {Michlmayr}}, \bibinfo {author} {\bibfnamefont {A.}~\bibnamefont {Vaterlaus}},\ and\ \bibinfo {author} {\bibfnamefont {Y.}~\bibnamefont {Acremann}},\ }\href {https://doi.org/10.1088/1361-648X/aa6a76} {\bibfield  {journal} {\bibinfo  {journal} {Journal of Physics: Condensed Matter}\ }\textbf {\bibinfo {volume} {29}},\ \bibinfo {pages} {214002} (\bibinfo {year} {2017})}\BibitemShut {NoStop}%
\bibitem [{\citenamefont {Jiménez-Cavero}\ \emph {et~al.}(2017)\citenamefont {Jiménez-Cavero}, \citenamefont {Lucas}, \citenamefont {Anadón}, \citenamefont {Ramos}, \citenamefont {Niizeki}, \citenamefont {Aguirre}, \citenamefont {Algarabel}, \citenamefont {Uchida}, \citenamefont {Ibarra}, \citenamefont {Saitoh},\ and\ \citenamefont {Morellón}}]{JimenezCavero2017}%
  \BibitemOpen
  \bibfield  {author} {\bibinfo {author} {\bibfnamefont {P.}~\bibnamefont {Jiménez-Cavero}}, \bibinfo {author} {\bibfnamefont {I.}~\bibnamefont {Lucas}}, \bibinfo {author} {\bibfnamefont {A.}~\bibnamefont {Anadón}}, \bibinfo {author} {\bibfnamefont {R.}~\bibnamefont {Ramos}}, \bibinfo {author} {\bibfnamefont {T.}~\bibnamefont {Niizeki}}, \bibinfo {author} {\bibfnamefont {M.~H.}\ \bibnamefont {Aguirre}}, \bibinfo {author} {\bibfnamefont {P.~A.}\ \bibnamefont {Algarabel}}, \bibinfo {author} {\bibfnamefont {K.}~\bibnamefont {Uchida}}, \bibinfo {author} {\bibfnamefont {M.~R.}\ \bibnamefont {Ibarra}}, \bibinfo {author} {\bibfnamefont {E.}~\bibnamefont {Saitoh}},\ and\ \bibinfo {author} {\bibfnamefont {L.}~\bibnamefont {Morellón}},\ }\href {https://doi.org/10.1063/1.4975618} {\bibfield  {journal} {\bibinfo  {journal} {APL Mater.}\ }\textbf {\bibinfo {volume} {5}},\ \bibinfo {pages} {026103} (\bibinfo {year} {2017})}\BibitemShut {NoStop}%
\bibitem [{\citenamefont {Winkelmann}\ \emph {et~al.}(2008)\citenamefont {Winkelmann}, \citenamefont {Hartung}, \citenamefont {Engelhard}, \citenamefont {Chiang},\ and\ \citenamefont {Kirschner}}]{Winkelmann2008}%
  \BibitemOpen
  \bibfield  {author} {\bibinfo {author} {\bibfnamefont {A.}~\bibnamefont {Winkelmann}}, \bibinfo {author} {\bibfnamefont {D.}~\bibnamefont {Hartung}}, \bibinfo {author} {\bibfnamefont {H.}~\bibnamefont {Engelhard}}, \bibinfo {author} {\bibfnamefont {C.-T.}\ \bibnamefont {Chiang}},\ and\ \bibinfo {author} {\bibfnamefont {J.}~\bibnamefont {Kirschner}},\ }\href {https://doi.org/10.1063/1.2949877} {\bibfield  {journal} {\bibinfo  {journal} {Rev. Sci. Instrum.}\ }\textbf {\bibinfo {volume} {79}},\ \bibinfo {pages} {083303} (\bibinfo {year} {2008})}\BibitemShut {NoStop}%
\bibitem [{\citenamefont {Andres}\ \emph {et~al.}(2015)\citenamefont {Andres}, \citenamefont {Christ}, \citenamefont {Gahl}, \citenamefont {Wietstruk}, \citenamefont {Weinelt},\ and\ \citenamefont {Kirschner}}]{Andres2015}%
  \BibitemOpen
  \bibfield  {author} {\bibinfo {author} {\bibfnamefont {B.}~\bibnamefont {Andres}}, \bibinfo {author} {\bibfnamefont {M.}~\bibnamefont {Christ}}, \bibinfo {author} {\bibfnamefont {C.}~\bibnamefont {Gahl}}, \bibinfo {author} {\bibfnamefont {M.}~\bibnamefont {Wietstruk}}, \bibinfo {author} {\bibfnamefont {M.}~\bibnamefont {Weinelt}},\ and\ \bibinfo {author} {\bibfnamefont {J.}~\bibnamefont {Kirschner}},\ }\href {https://doi.org/10.1103/PhysRevLett.115.207404} {\bibfield  {journal} {\bibinfo  {journal} {Phys. Rev. Lett.}\ }\textbf {\bibinfo {volume} {115}},\ \bibinfo {pages} {207404} (\bibinfo {year} {2015})}\BibitemShut {NoStop}%
\bibitem [{\citenamefont {Melnikov}\ \emph {et~al.}(2004)\citenamefont {Melnikov}, \citenamefont {Bovensiepen}, \citenamefont {Radu}, \citenamefont {Krupin}, \citenamefont {Starke}, \citenamefont {Matthias},\ and\ \citenamefont {Wolf}}]{Melnikov04}%
  \BibitemOpen
  \bibfield  {author} {\bibinfo {author} {\bibfnamefont {A.}~\bibnamefont {Melnikov}}, \bibinfo {author} {\bibfnamefont {U.}~\bibnamefont {Bovensiepen}}, \bibinfo {author} {\bibfnamefont {I.}~\bibnamefont {Radu}}, \bibinfo {author} {\bibfnamefont {O.}~\bibnamefont {Krupin}}, \bibinfo {author} {\bibfnamefont {K.}~\bibnamefont {Starke}}, \bibinfo {author} {\bibfnamefont {E.}~\bibnamefont {Matthias}},\ and\ \bibinfo {author} {\bibfnamefont {M.}~\bibnamefont {Wolf}},\ }\href {https://doi.org/https://doi.org/10.1016/j.jmmm.2003.12.359} {\bibfield  {journal} {\bibinfo  {journal} {Journal of Magnetism and Magnetic Materials}\ }\textbf {\bibinfo {volume} {272-276}},\ \bibinfo {pages} {1001} (\bibinfo {year} {2004})},\ \bibinfo {note} {proceedings of the International Conference on Magnetism (ICM 2003)}\BibitemShut {NoStop}%
\bibitem [{\citenamefont {Wu}\ \emph {et~al.}(2017)\citenamefont {Wu}, \citenamefont {Elyasi}, \citenamefont {Qiu}, \citenamefont {Chen}, \citenamefont {Liu}, \citenamefont {Ke},\ and\ \citenamefont {Yang}}]{Wu2017}%
  \BibitemOpen
  \bibfield  {author} {\bibinfo {author} {\bibfnamefont {Y.}~\bibnamefont {Wu}}, \bibinfo {author} {\bibfnamefont {M.}~\bibnamefont {Elyasi}}, \bibinfo {author} {\bibfnamefont {X.}~\bibnamefont {Qiu}}, \bibinfo {author} {\bibfnamefont {M.}~\bibnamefont {Chen}}, \bibinfo {author} {\bibfnamefont {Y.}~\bibnamefont {Liu}}, \bibinfo {author} {\bibfnamefont {L.}~\bibnamefont {Ke}},\ and\ \bibinfo {author} {\bibfnamefont {H.}~\bibnamefont {Yang}},\ }\href {https://doi.org/https://doi.org/10.1002/adma.201603031} {\bibfield  {journal} {\bibinfo  {journal} {Adv. Mater.}\ }\textbf {\bibinfo {volume} {29}},\ \bibinfo {pages} {1603031} (\bibinfo {year} {2017})}\BibitemShut {NoStop}%
\bibitem [{\citenamefont {Sasaki}\ \emph {et~al.}(2019)\citenamefont {Sasaki}, \citenamefont {Kota}, \citenamefont {Iihama}, \citenamefont {Suzuki}, \citenamefont {Sakuma},\ and\ \citenamefont {Mizukami}}]{Sasaki2019}%
  \BibitemOpen
  \bibfield  {author} {\bibinfo {author} {\bibfnamefont {Y.}~\bibnamefont {Sasaki}}, \bibinfo {author} {\bibfnamefont {Y.}~\bibnamefont {Kota}}, \bibinfo {author} {\bibfnamefont {S.}~\bibnamefont {Iihama}}, \bibinfo {author} {\bibfnamefont {K.~Z.}\ \bibnamefont {Suzuki}}, \bibinfo {author} {\bibfnamefont {A.}~\bibnamefont {Sakuma}},\ and\ \bibinfo {author} {\bibfnamefont {S.}~\bibnamefont {Mizukami}},\ }\href {https://doi.org/10.1103/PhysRevB.100.140406} {\bibfield  {journal} {\bibinfo  {journal} {Phys. Rev. B}\ }\textbf {\bibinfo {volume} {100}},\ \bibinfo {pages} {140406} (\bibinfo {year} {2019})}\BibitemShut {NoStop}%
\bibitem [{\citenamefont {Bierhance}\ \emph {et~al.}(2022)\citenamefont {Bierhance}, \citenamefont {Markou}, \citenamefont {Gueckstock}, \citenamefont {Rouzegar}, \citenamefont {Behovits}, \citenamefont {Chekhov}, \citenamefont {Wolf}, \citenamefont {Seifert}, \citenamefont {Felser},\ and\ \citenamefont {Kampfrath}}]{Bierhance2022}%
  \BibitemOpen
  \bibfield  {author} {\bibinfo {author} {\bibfnamefont {G.}~\bibnamefont {Bierhance}}, \bibinfo {author} {\bibfnamefont {A.}~\bibnamefont {Markou}}, \bibinfo {author} {\bibfnamefont {O.}~\bibnamefont {Gueckstock}}, \bibinfo {author} {\bibfnamefont {R.}~\bibnamefont {Rouzegar}}, \bibinfo {author} {\bibfnamefont {Y.}~\bibnamefont {Behovits}}, \bibinfo {author} {\bibfnamefont {A.~L.}\ \bibnamefont {Chekhov}}, \bibinfo {author} {\bibfnamefont {M.}~\bibnamefont {Wolf}}, \bibinfo {author} {\bibfnamefont {T.~S.}\ \bibnamefont {Seifert}}, \bibinfo {author} {\bibfnamefont {C.}~\bibnamefont {Felser}},\ and\ \bibinfo {author} {\bibfnamefont {T.}~\bibnamefont {Kampfrath}},\ }\href {https://doi.org/10.1063/5.0080308} {\bibfield  {journal} {\bibinfo  {journal} {Appl. Phys. Lett.}\ }\textbf {\bibinfo {volume} {120}},\ \bibinfo {pages} {082401} (\bibinfo {year} {2022})}\BibitemShut {NoStop}%
\bibitem [{\citenamefont {{Sasaki}}\ \emph {et~al.}(2020)\citenamefont {{Sasaki}}, \citenamefont {{Takahashi}},\ and\ \citenamefont {{Kasai}}}]{Sasaki2020}%
  \BibitemOpen
  \bibfield  {author} {\bibinfo {author} {\bibfnamefont {Y.}~\bibnamefont {{Sasaki}}}, \bibinfo {author} {\bibfnamefont {Y.}~\bibnamefont {{Takahashi}}},\ and\ \bibinfo {author} {\bibfnamefont {S.}~\bibnamefont {{Kasai}}},\ }\href {https://doi.org/10.35848/1882-0786/abb1c9} {\bibfield  {journal} {\bibinfo  {journal} {Appl. Phys. Express}\ }\textbf {\bibinfo {volume} {13}},\ \bibinfo {eid} {093003} (\bibinfo {year} {2020})}\BibitemShut {NoStop}%
\bibitem [{\citenamefont {Yao}\ \emph {et~al.}(2021)\citenamefont {Yao}, \citenamefont {Fu}, \citenamefont {Du}, \citenamefont {Huang}, \citenamefont {Lei}, \citenamefont {You},\ and\ \citenamefont {Xu}}]{Yao2021}%
  \BibitemOpen
  \bibfield  {author} {\bibinfo {author} {\bibfnamefont {Z.}~\bibnamefont {Yao}}, \bibinfo {author} {\bibfnamefont {H.}~\bibnamefont {Fu}}, \bibinfo {author} {\bibfnamefont {W.}~\bibnamefont {Du}}, \bibinfo {author} {\bibfnamefont {Y.}~\bibnamefont {Huang}}, \bibinfo {author} {\bibfnamefont {Z.}~\bibnamefont {Lei}}, \bibinfo {author} {\bibfnamefont {C.}~\bibnamefont {You}},\ and\ \bibinfo {author} {\bibfnamefont {X.}~\bibnamefont {Xu}},\ }\href {https://doi.org/10.1103/PhysRevB.103.L201404} {\bibfield  {journal} {\bibinfo  {journal} {Phys. Rev. B}\ }\textbf {\bibinfo {volume} {103}},\ \bibinfo {pages} {L201404} (\bibinfo {year} {2021})}\BibitemShut {NoStop}%
\bibitem [{\citenamefont {Heidtfeld}\ \emph {et~al.}(2021)\citenamefont {Heidtfeld}, \citenamefont {Adam}, \citenamefont {Kubota}, \citenamefont {Takanashi}, \citenamefont {Cao}, \citenamefont {Schmitz-Antoniak}, \citenamefont {B\"urgler}, \citenamefont {Wang}, \citenamefont {Greb}, \citenamefont {Chen}, \citenamefont {Komissarov}, \citenamefont {Hardtdegen}, \citenamefont {Mikulics}, \citenamefont {Sobolewski}, \citenamefont {Suga},\ and\ \citenamefont {Schneider}}]{Heidtfeld2021}%
  \BibitemOpen
  \bibfield  {author} {\bibinfo {author} {\bibfnamefont {S.}~\bibnamefont {Heidtfeld}}, \bibinfo {author} {\bibfnamefont {R.}~\bibnamefont {Adam}}, \bibinfo {author} {\bibfnamefont {T.}~\bibnamefont {Kubota}}, \bibinfo {author} {\bibfnamefont {K.}~\bibnamefont {Takanashi}}, \bibinfo {author} {\bibfnamefont {D.}~\bibnamefont {Cao}}, \bibinfo {author} {\bibfnamefont {C.}~\bibnamefont {Schmitz-Antoniak}}, \bibinfo {author} {\bibfnamefont {D.~E.}\ \bibnamefont {B\"urgler}}, \bibinfo {author} {\bibfnamefont {F.}~\bibnamefont {Wang}}, \bibinfo {author} {\bibfnamefont {C.}~\bibnamefont {Greb}}, \bibinfo {author} {\bibfnamefont {G.}~\bibnamefont {Chen}}, \bibinfo {author} {\bibfnamefont {I.}~\bibnamefont {Komissarov}}, \bibinfo {author} {\bibfnamefont {H.}~\bibnamefont {Hardtdegen}}, \bibinfo {author} {\bibfnamefont {M.}~\bibnamefont {Mikulics}}, \bibinfo {author} {\bibfnamefont {R.}~\bibnamefont {Sobolewski}}, \bibinfo {author} {\bibfnamefont {S.}~\bibnamefont {Suga}},\ and\ \bibinfo {author} {\bibfnamefont {C.~M.}\
  \bibnamefont {Schneider}},\ }\href {https://doi.org/10.1103/PhysRevResearch.3.043025} {\bibfield  {journal} {\bibinfo  {journal} {Phys. Rev. Res.}\ }\textbf {\bibinfo {volume} {3}},\ \bibinfo {pages} {043025} (\bibinfo {year} {2021})}\BibitemShut {NoStop}%
\bibitem [{\citenamefont {Gupta}\ \emph {et~al.}(2021)\citenamefont {Gupta}, \citenamefont {Husain}, \citenamefont {Kumar}, \citenamefont {Brucas}, \citenamefont {Rydberg},\ and\ \citenamefont {Svedlindh}}]{Gupta2021}%
  \BibitemOpen
  \bibfield  {author} {\bibinfo {author} {\bibfnamefont {R.}~\bibnamefont {Gupta}}, \bibinfo {author} {\bibfnamefont {S.}~\bibnamefont {Husain}}, \bibinfo {author} {\bibfnamefont {A.}~\bibnamefont {Kumar}}, \bibinfo {author} {\bibfnamefont {R.}~\bibnamefont {Brucas}}, \bibinfo {author} {\bibfnamefont {A.}~\bibnamefont {Rydberg}},\ and\ \bibinfo {author} {\bibfnamefont {P.}~\bibnamefont {Svedlindh}},\ }\href {https://doi.org/https://doi.org/10.1002/adom.202001987} {\bibfield  {journal} {\bibinfo  {journal} {Adv. Opt. Mater.}\ }\textbf {\bibinfo {volume} {9}},\ \bibinfo {pages} {2001987} (\bibinfo {year} {2021})}\BibitemShut {NoStop}%
\bibitem [{\citenamefont {Sch{\"o}nhense}\ and\ \citenamefont {Siegmann}(1993)}]{Schoenhense1993}%
  \BibitemOpen
  \bibfield  {author} {\bibinfo {author} {\bibfnamefont {G.}~\bibnamefont {Sch{\"o}nhense}}\ and\ \bibinfo {author} {\bibfnamefont {H.~C.}\ \bibnamefont {Siegmann}},\ }\href {https://doi.org/10.1002/andp.19935050504} {\bibfield  {journal} {\bibinfo  {journal} {Annalen der Physik}\ }\textbf {\bibinfo {volume} {2}},\ \bibinfo {pages} {465} (\bibinfo {year} {1993})}\BibitemShut {NoStop}%
\bibitem [{\citenamefont {Bovensiepen}(2007)}]{Bovensiepen2007}%
  \BibitemOpen
  \bibfield  {author} {\bibinfo {author} {\bibfnamefont {U.}~\bibnamefont {Bovensiepen}},\ }\href {http://stacks.iop.org/0953-8984/19/083201} {\bibfield  {journal} {\bibinfo  {journal} {J. Phys.: Condens. Matter}\ }\textbf {\bibinfo {volume} {19}},\ \bibinfo {pages} {083201} (\bibinfo {year} {2007})}\BibitemShut {NoStop}%
\bibitem [{\citenamefont {Knorren}\ \emph {et~al.}(2000)\citenamefont {Knorren}, \citenamefont {Bennemann}, \citenamefont {Burgermeister},\ and\ \citenamefont {Aeschlimann}}]{Knorren2000}%
  \BibitemOpen
  \bibfield  {author} {\bibinfo {author} {\bibfnamefont {R.}~\bibnamefont {Knorren}}, \bibinfo {author} {\bibfnamefont {K.~H.}\ \bibnamefont {Bennemann}}, \bibinfo {author} {\bibfnamefont {R.}~\bibnamefont {Burgermeister}},\ and\ \bibinfo {author} {\bibfnamefont {M.}~\bibnamefont {Aeschlimann}},\ }\href {https://doi.org/10.1103/PhysRevB.61.9427} {\bibfield  {journal} {\bibinfo  {journal} {Phys. Rev. B}\ }\textbf {\bibinfo {volume} {61}},\ \bibinfo {pages} {9427} (\bibinfo {year} {2000})}\BibitemShut {NoStop}%
\bibitem [{\citenamefont {Zhukov}\ \emph {et~al.}(2004)\citenamefont {Zhukov}, \citenamefont {Chulkov},\ and\ \citenamefont {Echenique}}]{Zhukov2004}%
  \BibitemOpen
  \bibfield  {author} {\bibinfo {author} {\bibfnamefont {V.~P.}\ \bibnamefont {Zhukov}}, \bibinfo {author} {\bibfnamefont {E.~V.}\ \bibnamefont {Chulkov}},\ and\ \bibinfo {author} {\bibfnamefont {P.~M.}\ \bibnamefont {Echenique}},\ }\href {https://doi.org/10.1103/PhysRevLett.93.096401} {\bibfield  {journal} {\bibinfo  {journal} {Phys. Rev. Lett.}\ }\textbf {\bibinfo {volume} {93}},\ \bibinfo {pages} {096401} (\bibinfo {year} {2004})}\BibitemShut {NoStop}%
\bibitem [{\citenamefont {Goris}\ \emph {et~al.}(2011)\citenamefont {Goris}, \citenamefont {Dobrich}, \citenamefont {Panzer}, \citenamefont {Schmidt}, \citenamefont {Donath},\ and\ \citenamefont {Weinelt}}]{Goris2011}%
  \BibitemOpen
  \bibfield  {author} {\bibinfo {author} {\bibfnamefont {A.}~\bibnamefont {Goris}}, \bibinfo {author} {\bibfnamefont {K.~M.}\ \bibnamefont {Dobrich}}, \bibinfo {author} {\bibfnamefont {I.}~\bibnamefont {Panzer}}, \bibinfo {author} {\bibfnamefont {A.~B.}\ \bibnamefont {Schmidt}}, \bibinfo {author} {\bibfnamefont {M.}~\bibnamefont {Donath}},\ and\ \bibinfo {author} {\bibfnamefont {M.}~\bibnamefont {Weinelt}},\ }\href {https://doi.org/10.1103/PhysRevLett.107.026601} {\bibfield  {journal} {\bibinfo  {journal} {Phys. Rev. Lett.}\ }\textbf {\bibinfo {volume} {107}},\ \bibinfo {pages} {026601} (\bibinfo {year} {2011})}\BibitemShut {NoStop}%
\bibitem [{\citenamefont {Kaltenborn}\ and\ \citenamefont {Schneider}(2014)}]{Kaltenborn2014}%
  \BibitemOpen
  \bibfield  {author} {\bibinfo {author} {\bibfnamefont {S.}~\bibnamefont {Kaltenborn}}\ and\ \bibinfo {author} {\bibfnamefont {H.~C.}\ \bibnamefont {Schneider}},\ }\href {https://doi.org/10.1103/PhysRevB.90.201104} {\bibfield  {journal} {\bibinfo  {journal} {Phys. Rev. B}\ }\textbf {\bibinfo {volume} {90}},\ \bibinfo {pages} {201104(R)} (\bibinfo {year} {2014})}\BibitemShut {NoStop}%
\bibitem [{\citenamefont {Tengdin}\ \emph {et~al.}(2018)\citenamefont {Tengdin}, \citenamefont {You}, \citenamefont {Chen}, \citenamefont {Shi}, \citenamefont {Zusin}, \citenamefont {Zhang}, \citenamefont {Gentry}, \citenamefont {Blonsky}, \citenamefont {Keller}, \citenamefont {Oppeneer}, \citenamefont {Kapteyn}, \citenamefont {Tao},\ and\ \citenamefont {Murnane}}]{Tengdin2018}%
  \BibitemOpen
  \bibfield  {author} {\bibinfo {author} {\bibfnamefont {P.}~\bibnamefont {Tengdin}}, \bibinfo {author} {\bibfnamefont {W.}~\bibnamefont {You}}, \bibinfo {author} {\bibfnamefont {C.}~\bibnamefont {Chen}}, \bibinfo {author} {\bibfnamefont {X.}~\bibnamefont {Shi}}, \bibinfo {author} {\bibfnamefont {D.}~\bibnamefont {Zusin}}, \bibinfo {author} {\bibfnamefont {Y.}~\bibnamefont {Zhang}}, \bibinfo {author} {\bibfnamefont {C.}~\bibnamefont {Gentry}}, \bibinfo {author} {\bibfnamefont {A.}~\bibnamefont {Blonsky}}, \bibinfo {author} {\bibfnamefont {M.}~\bibnamefont {Keller}}, \bibinfo {author} {\bibfnamefont {P.~M.}\ \bibnamefont {Oppeneer}}, \bibinfo {author} {\bibfnamefont {H.~C.}\ \bibnamefont {Kapteyn}}, \bibinfo {author} {\bibfnamefont {Z.}~\bibnamefont {Tao}},\ and\ \bibinfo {author} {\bibfnamefont {M.~M.}\ \bibnamefont {Murnane}},\ }\href {https://doi.org/10.1126/sciadv.aap9744} {\bibfield  {journal} {\bibinfo  {journal} {Science Advances}\ }\textbf {\bibinfo {volume} {4}},\ \bibinfo {pages} {eaap9744} (\bibinfo
  {year} {2018})}\BibitemShut {NoStop}%
\bibitem [{\citenamefont {Kirschner}(1985)}]{Kirschner1985}%
  \BibitemOpen
  \bibfield  {author} {\bibinfo {author} {\bibfnamefont {J.}~\bibnamefont {Kirschner}},\ }\href@noop {} {\bibfield  {journal} {\bibinfo  {journal} {Phys. Rev. Lett.}\ }\textbf {\bibinfo {volume} {55}},\ \bibinfo {pages} {973} (\bibinfo {year} {1985})}\BibitemShut {NoStop}%
\bibitem [{\citenamefont {Bode}\ \emph {et~al.}(1999)\citenamefont {Bode}, \citenamefont {Getzlaff}, \citenamefont {Kubetzka}, \citenamefont {Pascal}, \citenamefont {Pietzsch},\ and\ \citenamefont {Wiesendanger}}]{Bode1999}%
  \BibitemOpen
  \bibfield  {author} {\bibinfo {author} {\bibfnamefont {M.}~\bibnamefont {Bode}}, \bibinfo {author} {\bibfnamefont {M.}~\bibnamefont {Getzlaff}}, \bibinfo {author} {\bibfnamefont {A.}~\bibnamefont {Kubetzka}}, \bibinfo {author} {\bibfnamefont {R.}~\bibnamefont {Pascal}}, \bibinfo {author} {\bibfnamefont {O.}~\bibnamefont {Pietzsch}},\ and\ \bibinfo {author} {\bibfnamefont {R.}~\bibnamefont {Wiesendanger}},\ }\href {https://doi.org/10.1103/PhysRevLett.83.3017} {\bibfield  {journal} {\bibinfo  {journal} {Phys. Rev. Lett.}\ }\textbf {\bibinfo {volume} {83}},\ \bibinfo {pages} {3017} (\bibinfo {year} {1999})}\BibitemShut {NoStop}%
\bibitem [{\citenamefont {Bobowski}\ \emph {et~al.}(2024)\citenamefont {Bobowski}, \citenamefont {Zheng}, \citenamefont {Frietsch}, \citenamefont {Lawrenz}, \citenamefont {Bronsch}, \citenamefont {Gahl}, \citenamefont {Andres}, \citenamefont {Str\"uber}, \citenamefont {Weinelt}, \citenamefont {Carley}, \citenamefont {Teichmann}, \citenamefont {Scherz}, \citenamefont {Molodtsov}, \citenamefont {Cacho}, \citenamefont {Chapman},\ and\ \citenamefont {Springate}}]{Bobowski2024}%
  \BibitemOpen
  \bibfield  {author} {\bibinfo {author} {\bibfnamefont {K.}~\bibnamefont {Bobowski}}, \bibinfo {author} {\bibfnamefont {X.}~\bibnamefont {Zheng}}, \bibinfo {author} {\bibfnamefont {B.}~\bibnamefont {Frietsch}}, \bibinfo {author} {\bibfnamefont {D.}~\bibnamefont {Lawrenz}}, \bibinfo {author} {\bibfnamefont {W.}~\bibnamefont {Bronsch}}, \bibinfo {author} {\bibfnamefont {C.}~\bibnamefont {Gahl}}, \bibinfo {author} {\bibfnamefont {B.}~\bibnamefont {Andres}}, \bibinfo {author} {\bibfnamefont {C.}~\bibnamefont {Str\"uber}}, \bibinfo {author} {\bibfnamefont {M.}~\bibnamefont {Weinelt}}, \bibinfo {author} {\bibfnamefont {R.}~\bibnamefont {Carley}}, \bibinfo {author} {\bibfnamefont {M.}~\bibnamefont {Teichmann}}, \bibinfo {author} {\bibfnamefont {A.}~\bibnamefont {Scherz}}, \bibinfo {author} {\bibfnamefont {S.}~\bibnamefont {Molodtsov}}, \bibinfo {author} {\bibfnamefont {C.}~\bibnamefont {Cacho}}, \bibinfo {author} {\bibfnamefont {R.~T.}\ \bibnamefont {Chapman}},\ and\ \bibinfo {author} {\bibfnamefont
  {E.}~\bibnamefont {Springate}},\ }\href@noop {} {\bibfield  {journal} {\bibinfo  {journal} {in press}\ } (\bibinfo {year} {2024})}\BibitemShut {NoStop}%
\bibitem [{\citenamefont {Thielemann-Kühn}\ \emph {et~al.}(2024)\citenamefont {Thielemann-Kühn}, \citenamefont {Amrhein}, \citenamefont {Bronsch}, \citenamefont {Jana}, \citenamefont {Pontius}, \citenamefont {Engel}, \citenamefont {Miedema}, \citenamefont {Legut}, \citenamefont {Carva}, \citenamefont {Atxitia}, \citenamefont {van Kuiken}, \citenamefont {Teichmann}, \citenamefont {Carley}, \citenamefont {Mercadier}, \citenamefont {Yaroslavtsev}, \citenamefont {Mercurio}, \citenamefont {Guyader}, \citenamefont {Agarwal}, \citenamefont {Gort}, \citenamefont {Scherz}, \citenamefont {Dziarzhytski}, \citenamefont {Brenner}, \citenamefont {Pressacco}, \citenamefont {Wang}, \citenamefont {Schunck}, \citenamefont {Sinha}, \citenamefont {Beye}, \citenamefont {Chiuzbăian}, \citenamefont {Oppeneer}, \citenamefont {Weinelt},\ and\ \citenamefont {Schüßler-Langeheine}}]{Thielemann2024}%
  \BibitemOpen
  \bibfield  {author} {\bibinfo {author} {\bibfnamefont {N.}~\bibnamefont {Thielemann-Kühn}}, \bibinfo {author} {\bibfnamefont {T.}~\bibnamefont {Amrhein}}, \bibinfo {author} {\bibfnamefont {W.}~\bibnamefont {Bronsch}}, \bibinfo {author} {\bibfnamefont {S.}~\bibnamefont {Jana}}, \bibinfo {author} {\bibfnamefont {N.}~\bibnamefont {Pontius}}, \bibinfo {author} {\bibfnamefont {R.~Y.}\ \bibnamefont {Engel}}, \bibinfo {author} {\bibfnamefont {P.~S.}\ \bibnamefont {Miedema}}, \bibinfo {author} {\bibfnamefont {D.}~\bibnamefont {Legut}}, \bibinfo {author} {\bibfnamefont {K.}~\bibnamefont {Carva}}, \bibinfo {author} {\bibfnamefont {U.}~\bibnamefont {Atxitia}}, \bibinfo {author} {\bibfnamefont {B.~E.}\ \bibnamefont {van Kuiken}}, \bibinfo {author} {\bibfnamefont {M.}~\bibnamefont {Teichmann}}, \bibinfo {author} {\bibfnamefont {R.~E.}\ \bibnamefont {Carley}}, \bibinfo {author} {\bibfnamefont {L.}~\bibnamefont {Mercadier}}, \bibinfo {author} {\bibfnamefont {A.}~\bibnamefont {Yaroslavtsev}}, \bibinfo {author}
  {\bibfnamefont {G.}~\bibnamefont {Mercurio}}, \bibinfo {author} {\bibfnamefont {L.~L.}\ \bibnamefont {Guyader}}, \bibinfo {author} {\bibfnamefont {N.}~\bibnamefont {Agarwal}}, \bibinfo {author} {\bibfnamefont {R.}~\bibnamefont {Gort}}, \bibinfo {author} {\bibfnamefont {A.}~\bibnamefont {Scherz}}, \bibinfo {author} {\bibfnamefont {S.}~\bibnamefont {Dziarzhytski}}, \bibinfo {author} {\bibfnamefont {G.}~\bibnamefont {Brenner}}, \bibinfo {author} {\bibfnamefont {F.}~\bibnamefont {Pressacco}}, \bibinfo {author} {\bibfnamefont {R.-P.}\ \bibnamefont {Wang}}, \bibinfo {author} {\bibfnamefont {J.~O.}\ \bibnamefont {Schunck}}, \bibinfo {author} {\bibfnamefont {M.}~\bibnamefont {Sinha}}, \bibinfo {author} {\bibfnamefont {M.}~\bibnamefont {Beye}}, \bibinfo {author} {\bibfnamefont {G.~S.}\ \bibnamefont {Chiuzbăian}}, \bibinfo {author} {\bibfnamefont {P.~M.}\ \bibnamefont {Oppeneer}}, \bibinfo {author} {\bibfnamefont {M.}~\bibnamefont {Weinelt}},\ and\ \bibinfo {author} {\bibfnamefont {C.}~\bibnamefont
  {Schüßler-Langeheine}},\ }\href {https://doi.org/10.1126/sciadv.adk9522} {\bibfield  {journal} {\bibinfo  {journal} {Science Advances}\ }\textbf {\bibinfo {volume} {10}},\ \bibinfo {pages} {eadk9522} (\bibinfo {year} {2024})}\BibitemShut {NoStop}%
\bibitem [{\citenamefont {Roth}\ \emph {et~al.}(2012)\citenamefont {Roth}, \citenamefont {Schellekens}, \citenamefont {Alebrand}, \citenamefont {Schmitt}, \citenamefont {Steil}, \citenamefont {Koopmans}, \citenamefont {Cinchetti},\ and\ \citenamefont {Aeschlimann}}]{Roth2012}%
  \BibitemOpen
  \bibfield  {author} {\bibinfo {author} {\bibfnamefont {T.}~\bibnamefont {Roth}}, \bibinfo {author} {\bibfnamefont {A.~J.}\ \bibnamefont {Schellekens}}, \bibinfo {author} {\bibfnamefont {S.}~\bibnamefont {Alebrand}}, \bibinfo {author} {\bibfnamefont {O.}~\bibnamefont {Schmitt}}, \bibinfo {author} {\bibfnamefont {D.}~\bibnamefont {Steil}}, \bibinfo {author} {\bibfnamefont {B.}~\bibnamefont {Koopmans}}, \bibinfo {author} {\bibfnamefont {M.}~\bibnamefont {Cinchetti}},\ and\ \bibinfo {author} {\bibfnamefont {M.}~\bibnamefont {Aeschlimann}},\ }\href {https://doi.org/10.1103/PhysRevX.2.021006} {\bibfield  {journal} {\bibinfo  {journal} {Phys. Rev. X}\ }\textbf {\bibinfo {volume} {2}},\ \bibinfo {pages} {021006} (\bibinfo {year} {2012})}\BibitemShut {NoStop}%
\bibitem [{\citenamefont {Rhie}\ \emph {et~al.}(2003)\citenamefont {Rhie}, \citenamefont {D\"urr},\ and\ \citenamefont {Eberhardt}}]{Rhie2003}%
  \BibitemOpen
  \bibfield  {author} {\bibinfo {author} {\bibfnamefont {H.-S.}\ \bibnamefont {Rhie}}, \bibinfo {author} {\bibfnamefont {H.~A.}\ \bibnamefont {D\"urr}},\ and\ \bibinfo {author} {\bibfnamefont {W.}~\bibnamefont {Eberhardt}},\ }\href {https://doi.org/10.1103/PhysRevLett.90.247201} {\bibfield  {journal} {\bibinfo  {journal} {Phys. Rev. Lett.}\ }\textbf {\bibinfo {volume} {90}},\ \bibinfo {pages} {247201} (\bibinfo {year} {2003})}\BibitemShut {NoStop}%
\bibitem [{\citenamefont {Stamm}\ \emph {et~al.}(2007)\citenamefont {Stamm}, \citenamefont {Kachel}, \citenamefont {Pontius}, \citenamefont {Mitzner}, \citenamefont {Quast}, \citenamefont {Holldack}, \citenamefont {Khan}, \citenamefont {Lupulescu}, \citenamefont {Aziz}, \citenamefont {Wietstruk}, \citenamefont {D{\"u}rr},\ and\ \citenamefont {Eberhardt}}]{Stamm2007}%
  \BibitemOpen
  \bibfield  {author} {\bibinfo {author} {\bibfnamefont {C.}~\bibnamefont {Stamm}}, \bibinfo {author} {\bibfnamefont {T.}~\bibnamefont {Kachel}}, \bibinfo {author} {\bibfnamefont {N.}~\bibnamefont {Pontius}}, \bibinfo {author} {\bibfnamefont {R.}~\bibnamefont {Mitzner}}, \bibinfo {author} {\bibfnamefont {T.}~\bibnamefont {Quast}}, \bibinfo {author} {\bibfnamefont {K.}~\bibnamefont {Holldack}}, \bibinfo {author} {\bibfnamefont {S.}~\bibnamefont {Khan}}, \bibinfo {author} {\bibfnamefont {C.}~\bibnamefont {Lupulescu}}, \bibinfo {author} {\bibfnamefont {E.~F.}\ \bibnamefont {Aziz}}, \bibinfo {author} {\bibfnamefont {M.}~\bibnamefont {Wietstruk}}, \bibinfo {author} {\bibfnamefont {H.~A.}\ \bibnamefont {D{\"u}rr}},\ and\ \bibinfo {author} {\bibfnamefont {W.}~\bibnamefont {Eberhardt}},\ }\href {https://doi.org/10.1038/nmat1985} {\bibfield  {journal} {\bibinfo  {journal} {Nat. Mater.}\ }\textbf {\bibinfo {volume} {6}},\ \bibinfo {pages} {740} (\bibinfo {year} {2007})}\BibitemShut {NoStop}%
\bibitem [{\citenamefont {Krau\ss{}}\ \emph {et~al.}(2009)\citenamefont {Krau\ss{}}, \citenamefont {Roth}, \citenamefont {Alebrand}, \citenamefont {Steil}, \citenamefont {Cinchetti}, \citenamefont {Aeschlimann},\ and\ \citenamefont {Schneider}}]{Krauss2009}%
  \BibitemOpen
  \bibfield  {author} {\bibinfo {author} {\bibfnamefont {M.}~\bibnamefont {Krau\ss{}}}, \bibinfo {author} {\bibfnamefont {T.}~\bibnamefont {Roth}}, \bibinfo {author} {\bibfnamefont {S.}~\bibnamefont {Alebrand}}, \bibinfo {author} {\bibfnamefont {D.}~\bibnamefont {Steil}}, \bibinfo {author} {\bibfnamefont {M.}~\bibnamefont {Cinchetti}}, \bibinfo {author} {\bibfnamefont {M.}~\bibnamefont {Aeschlimann}},\ and\ \bibinfo {author} {\bibfnamefont {H.~C.}\ \bibnamefont {Schneider}},\ }\href {https://doi.org/10.1103/PhysRevB.80.180407} {\bibfield  {journal} {\bibinfo  {journal} {Phys. Rev. B}\ }\textbf {\bibinfo {volume} {80}},\ \bibinfo {pages} {180407} (\bibinfo {year} {2009})}\BibitemShut {NoStop}%
\bibitem [{\citenamefont {Kuiper}\ \emph {et~al.}(2014)\citenamefont {Kuiper}, \citenamefont {Roth}, \citenamefont {Schellekens}, \citenamefont {Schmitt}, \citenamefont {Koopmans}, \citenamefont {Cinchetti},\ and\ \citenamefont {Aeschlimann}}]{Kupier2014}%
  \BibitemOpen
  \bibfield  {author} {\bibinfo {author} {\bibfnamefont {K.~C.}\ \bibnamefont {Kuiper}}, \bibinfo {author} {\bibfnamefont {T.}~\bibnamefont {Roth}}, \bibinfo {author} {\bibfnamefont {A.~J.}\ \bibnamefont {Schellekens}}, \bibinfo {author} {\bibfnamefont {O.}~\bibnamefont {Schmitt}}, \bibinfo {author} {\bibfnamefont {B.}~\bibnamefont {Koopmans}}, \bibinfo {author} {\bibfnamefont {M.}~\bibnamefont {Cinchetti}},\ and\ \bibinfo {author} {\bibfnamefont {M.}~\bibnamefont {Aeschlimann}},\ }\href {https://doi.org/10.1063/1.4902069} {\bibfield  {journal} {\bibinfo  {journal} {Applied Physics Letters}\ }\textbf {\bibinfo {volume} {105}},\ \bibinfo {pages} {202402} (\bibinfo {year} {2014})}\BibitemShut {NoStop}%
\bibitem [{\citenamefont {Cinchetti}\ \emph {et~al.}(2006)\citenamefont {Cinchetti}, \citenamefont {S\'anchez~Albaneda}, \citenamefont {Hoffmann}, \citenamefont {Roth}, \citenamefont {W\"ustenberg}, \citenamefont {Krau\ss{}}, \citenamefont {Andreyev}, \citenamefont {Schneider}, \citenamefont {Bauer},\ and\ \citenamefont {Aeschlimann}}]{Cinchetti2006}%
  \BibitemOpen
  \bibfield  {author} {\bibinfo {author} {\bibfnamefont {M.}~\bibnamefont {Cinchetti}}, \bibinfo {author} {\bibfnamefont {M.}~\bibnamefont {S\'anchez~Albaneda}}, \bibinfo {author} {\bibfnamefont {D.}~\bibnamefont {Hoffmann}}, \bibinfo {author} {\bibfnamefont {T.}~\bibnamefont {Roth}}, \bibinfo {author} {\bibfnamefont {J.-P.}\ \bibnamefont {W\"ustenberg}}, \bibinfo {author} {\bibfnamefont {M.}~\bibnamefont {Krau\ss{}}}, \bibinfo {author} {\bibfnamefont {O.}~\bibnamefont {Andreyev}}, \bibinfo {author} {\bibfnamefont {H.~C.}\ \bibnamefont {Schneider}}, \bibinfo {author} {\bibfnamefont {M.}~\bibnamefont {Bauer}},\ and\ \bibinfo {author} {\bibfnamefont {M.}~\bibnamefont {Aeschlimann}},\ }\href {https://doi.org/10.1103/PhysRevLett.97.177201} {\bibfield  {journal} {\bibinfo  {journal} {Phys. Rev. Lett.}\ }\textbf {\bibinfo {volume} {97}},\ \bibinfo {pages} {177201} (\bibinfo {year} {2006})}\BibitemShut {NoStop}%
\bibitem [{\citenamefont {Weber}\ \emph {et~al.}(2011)\citenamefont {Weber}, \citenamefont {Pressacco}, \citenamefont {G\"unther}, \citenamefont {Mancini}, \citenamefont {Oppeneer},\ and\ \citenamefont {Back}}]{Weber2011}%
  \BibitemOpen
  \bibfield  {author} {\bibinfo {author} {\bibfnamefont {A.}~\bibnamefont {Weber}}, \bibinfo {author} {\bibfnamefont {F.}~\bibnamefont {Pressacco}}, \bibinfo {author} {\bibfnamefont {S.}~\bibnamefont {G\"unther}}, \bibinfo {author} {\bibfnamefont {E.}~\bibnamefont {Mancini}}, \bibinfo {author} {\bibfnamefont {P.~M.}\ \bibnamefont {Oppeneer}},\ and\ \bibinfo {author} {\bibfnamefont {C.~H.}\ \bibnamefont {Back}},\ }\href {https://doi.org/10.1103/PhysRevB.84.132412} {\bibfield  {journal} {\bibinfo  {journal} {Phys. Rev. B}\ }\textbf {\bibinfo {volume} {84}},\ \bibinfo {pages} {132412} (\bibinfo {year} {2011})}\BibitemShut {NoStop}%
\bibitem [{\citenamefont {Carpene}\ \emph {et~al.}(2008)\citenamefont {Carpene}, \citenamefont {Mancini}, \citenamefont {Dallera}, \citenamefont {Brenna}, \citenamefont {Puppin},\ and\ \citenamefont {De~Silvestri}}]{Carpene2008}%
  \BibitemOpen
  \bibfield  {author} {\bibinfo {author} {\bibfnamefont {E.}~\bibnamefont {Carpene}}, \bibinfo {author} {\bibfnamefont {E.}~\bibnamefont {Mancini}}, \bibinfo {author} {\bibfnamefont {C.}~\bibnamefont {Dallera}}, \bibinfo {author} {\bibfnamefont {M.}~\bibnamefont {Brenna}}, \bibinfo {author} {\bibfnamefont {E.}~\bibnamefont {Puppin}},\ and\ \bibinfo {author} {\bibfnamefont {S.}~\bibnamefont {De~Silvestri}},\ }\href {https://doi.org/10.1103/PhysRevB.78.174422} {\bibfield  {journal} {\bibinfo  {journal} {Phys. Rev. B}\ }\textbf {\bibinfo {volume} {78}},\ \bibinfo {pages} {174422} (\bibinfo {year} {2008})}\BibitemShut {NoStop}%
\bibitem [{\citenamefont {Jensen}\ and\ \citenamefont {Mackintosh}(1991)}]{Jensen1991}%
  \BibitemOpen
  \bibfield  {author} {\bibinfo {author} {\bibfnamefont {J.}~\bibnamefont {Jensen}}\ and\ \bibinfo {author} {\bibfnamefont {A.~R.}\ \bibnamefont {Mackintosh}},\ }in\ \href {https://doi.org/10.1093/oso/9780198520276.003.0002} {\emph {\bibinfo {booktitle} {Rare Earth Magnetism: Structures and Excitations}}}\ (\bibinfo  {publisher} {Oxford University Press},\ \bibinfo {year} {1991})\BibitemShut {NoStop}%
\bibitem [{\citenamefont {Stanciu}\ \emph {et~al.}(2007)\citenamefont {Stanciu}, \citenamefont {Hansteen}, \citenamefont {Kimel}, \citenamefont {Kirilyuk}, \citenamefont {Tsukamoto}, \citenamefont {Itoh},\ and\ \citenamefont {Rasing}}]{Stanciu2007}%
  \BibitemOpen
  \bibfield  {author} {\bibinfo {author} {\bibfnamefont {C.~D.}\ \bibnamefont {Stanciu}}, \bibinfo {author} {\bibfnamefont {F.}~\bibnamefont {Hansteen}}, \bibinfo {author} {\bibfnamefont {A.~V.}\ \bibnamefont {Kimel}}, \bibinfo {author} {\bibfnamefont {A.}~\bibnamefont {Kirilyuk}}, \bibinfo {author} {\bibfnamefont {A.}~\bibnamefont {Tsukamoto}}, \bibinfo {author} {\bibfnamefont {A.}~\bibnamefont {Itoh}},\ and\ \bibinfo {author} {\bibfnamefont {T.}~\bibnamefont {Rasing}},\ }\href@noop {} {\bibfield  {journal} {\bibinfo  {journal} {Phys. Rev. Lett.}\ }\textbf {\bibinfo {volume} {99}},\ \bibinfo {pages} {047601} (\bibinfo {year} {2007})}\BibitemShut {NoStop}%
\bibitem [{\citenamefont {Ostler}\ \emph {et~al.}(2012)\citenamefont {Ostler}, \citenamefont {Barker}, \citenamefont {Evans}, \citenamefont {Chantrell}, \citenamefont {Atxitia}, \citenamefont {Chubykalo-Fesenko}, \citenamefont {El~Moussaoui}, \citenamefont {Le~Guyader}, \citenamefont {Mengotti}, \citenamefont {Heyderman}, \citenamefont {Nolting}, \citenamefont {Tsukamoto}, \citenamefont {Itoh}, \citenamefont {Afanasiev}, \citenamefont {Ivanov}, \citenamefont {Kalashnikova}, \citenamefont {Vahaplar}, \citenamefont {Mentink}, \citenamefont {Kirilyuk}, \citenamefont {Rasing},\ and\ \citenamefont {Kimel}}]{Ostler2012}%
  \BibitemOpen
  \bibfield  {author} {\bibinfo {author} {\bibfnamefont {T.}~\bibnamefont {Ostler}}, \bibinfo {author} {\bibfnamefont {J.}~\bibnamefont {Barker}}, \bibinfo {author} {\bibfnamefont {R.}~\bibnamefont {Evans}}, \bibinfo {author} {\bibfnamefont {R.}~\bibnamefont {Chantrell}}, \bibinfo {author} {\bibfnamefont {U.}~\bibnamefont {Atxitia}}, \bibinfo {author} {\bibfnamefont {O.}~\bibnamefont {Chubykalo-Fesenko}}, \bibinfo {author} {\bibfnamefont {S.}~\bibnamefont {El~Moussaoui}}, \bibinfo {author} {\bibfnamefont {L.}~\bibnamefont {Le~Guyader}}, \bibinfo {author} {\bibfnamefont {E.}~\bibnamefont {Mengotti}}, \bibinfo {author} {\bibfnamefont {L.}~\bibnamefont {Heyderman}}, \bibinfo {author} {\bibfnamefont {F.}~\bibnamefont {Nolting}}, \bibinfo {author} {\bibfnamefont {A.}~\bibnamefont {Tsukamoto}}, \bibinfo {author} {\bibfnamefont {A.}~\bibnamefont {Itoh}}, \bibinfo {author} {\bibfnamefont {D.}~\bibnamefont {Afanasiev}}, \bibinfo {author} {\bibfnamefont {B.}~\bibnamefont {Ivanov}}, \bibinfo {author} {\bibfnamefont
  {A.}~\bibnamefont {Kalashnikova}}, \bibinfo {author} {\bibfnamefont {K.}~\bibnamefont {Vahaplar}}, \bibinfo {author} {\bibfnamefont {J.}~\bibnamefont {Mentink}}, \bibinfo {author} {\bibfnamefont {A.}~\bibnamefont {Kirilyuk}}, \bibinfo {author} {\bibfnamefont {T.}~\bibnamefont {Rasing}},\ and\ \bibinfo {author} {\bibfnamefont {A.}~\bibnamefont {Kimel}},\ }\href {https://doi.org/10.1038/ncomms1666} {\bibfield  {journal} {\bibinfo  {journal} {Nat. Commun.}\ }\textbf {\bibinfo {volume} {3}},\ \bibinfo {pages} {666} (\bibinfo {year} {2012})}\BibitemShut {NoStop}%
\bibitem [{\citenamefont {Kirilyuk}\ \emph {et~al.}(2013)\citenamefont {Kirilyuk}, \citenamefont {Kimel},\ and\ \citenamefont {Rasing}}]{Kirilyuk2013}%
  \BibitemOpen
  \bibfield  {author} {\bibinfo {author} {\bibfnamefont {A.}~\bibnamefont {Kirilyuk}}, \bibinfo {author} {\bibfnamefont {A.~V.}\ \bibnamefont {Kimel}},\ and\ \bibinfo {author} {\bibfnamefont {T.}~\bibnamefont {Rasing}},\ }\href {http://stacks.iop.org/0034-4885/76/i=2/a=026501} {\bibfield  {journal} {\bibinfo  {journal} {Reports on Progress in Physics}\ }\textbf {\bibinfo {volume} {76}},\ \bibinfo {pages} {026501} (\bibinfo {year} {2013})}\BibitemShut {NoStop}%
\bibitem [{\citenamefont {Graves}\ \emph {et~al.}(2013)\citenamefont {Graves}, \citenamefont {Reid}, \citenamefont {Wang}, \citenamefont {Wu}, \citenamefont {de~Jong}, \citenamefont {Vahaplar}, \citenamefont {Radu}, \citenamefont {Bernstein}, \citenamefont {Messerschmidt}, \citenamefont {M\"{u}ller}, \citenamefont {Coffee}, \citenamefont {Bionta}, \citenamefont {Epp}, \citenamefont {Hartmann}, \citenamefont {Kimmel}, \citenamefont {Hauser}, \citenamefont {Hartmann}, \citenamefont {Holl}, \citenamefont {Gorke}, \citenamefont {Mentink}, \citenamefont {Tsukamoto}, \citenamefont {Fognini}, \citenamefont {Turner}, \citenamefont {Schlotter}, \citenamefont {Rolles}, \citenamefont {Soltau}, \citenamefont {Str\"{u}der}, \citenamefont {Acremann}, \citenamefont {Kimel}, \citenamefont {Kirilyuk}, \citenamefont {Rasing}, \citenamefont {St\"{o}hr}, \citenamefont {Scherz},\ and\ \citenamefont {D\"{u}rr}}]{Graves2013}%
  \BibitemOpen
  \bibfield  {author} {\bibinfo {author} {\bibfnamefont {C.~E.}\ \bibnamefont {Graves}}, \bibinfo {author} {\bibfnamefont {A.~H.}\ \bibnamefont {Reid}}, \bibinfo {author} {\bibfnamefont {T.}~\bibnamefont {Wang}}, \bibinfo {author} {\bibfnamefont {B.}~\bibnamefont {Wu}}, \bibinfo {author} {\bibfnamefont {S.}~\bibnamefont {de~Jong}}, \bibinfo {author} {\bibfnamefont {K.}~\bibnamefont {Vahaplar}}, \bibinfo {author} {\bibfnamefont {I.}~\bibnamefont {Radu}}, \bibinfo {author} {\bibfnamefont {D.~P.}\ \bibnamefont {Bernstein}}, \bibinfo {author} {\bibfnamefont {M.}~\bibnamefont {Messerschmidt}}, \bibinfo {author} {\bibfnamefont {L.}~\bibnamefont {M\"{u}ller}}, \bibinfo {author} {\bibfnamefont {R.}~\bibnamefont {Coffee}}, \bibinfo {author} {\bibfnamefont {M.}~\bibnamefont {Bionta}}, \bibinfo {author} {\bibfnamefont {S.~W.}\ \bibnamefont {Epp}}, \bibinfo {author} {\bibfnamefont {R.}~\bibnamefont {Hartmann}}, \bibinfo {author} {\bibfnamefont {N.}~\bibnamefont {Kimmel}}, \bibinfo {author} {\bibfnamefont
  {G.}~\bibnamefont {Hauser}}, \bibinfo {author} {\bibfnamefont {A.}~\bibnamefont {Hartmann}}, \bibinfo {author} {\bibfnamefont {P.}~\bibnamefont {Holl}}, \bibinfo {author} {\bibfnamefont {H.}~\bibnamefont {Gorke}}, \bibinfo {author} {\bibfnamefont {J.~H.}\ \bibnamefont {Mentink}}, \bibinfo {author} {\bibfnamefont {A.}~\bibnamefont {Tsukamoto}}, \bibinfo {author} {\bibfnamefont {A.}~\bibnamefont {Fognini}}, \bibinfo {author} {\bibfnamefont {J.~J.}\ \bibnamefont {Turner}}, \bibinfo {author} {\bibfnamefont {W.~F.}\ \bibnamefont {Schlotter}}, \bibinfo {author} {\bibfnamefont {D.}~\bibnamefont {Rolles}}, \bibinfo {author} {\bibfnamefont {H.}~\bibnamefont {Soltau}}, \bibinfo {author} {\bibfnamefont {L.}~\bibnamefont {Str\"{u}der}}, \bibinfo {author} {\bibfnamefont {Y.}~\bibnamefont {Acremann}}, \bibinfo {author} {\bibfnamefont {A.~V.}\ \bibnamefont {Kimel}}, \bibinfo {author} {\bibfnamefont {A.}~\bibnamefont {Kirilyuk}}, \bibinfo {author} {\bibfnamefont {T.}~\bibnamefont {Rasing}}, \bibinfo {author} {\bibfnamefont
  {J.}~\bibnamefont {St\"{o}hr}}, \bibinfo {author} {\bibfnamefont {A.~O.}\ \bibnamefont {Scherz}},\ and\ \bibinfo {author} {\bibfnamefont {H.~A.}\ \bibnamefont {D\"{u}rr}},\ }\href {https://doi.org/10.1038/nmat3597} {\bibfield  {journal} {\bibinfo  {journal} {Nat. Mater.}\ }\textbf {\bibinfo {volume} {12}},\ \bibinfo {pages} {293} (\bibinfo {year} {2013})}\BibitemShut {NoStop}%
\bibitem [{\citenamefont {Beens}\ \emph {et~al.}(2019)\citenamefont {Beens}, \citenamefont {Lalieu}, \citenamefont {Deenen}, \citenamefont {Duine},\ and\ \citenamefont {Koopmans}}]{Beens2019}%
  \BibitemOpen
  \bibfield  {author} {\bibinfo {author} {\bibfnamefont {M.}~\bibnamefont {Beens}}, \bibinfo {author} {\bibfnamefont {M.~L.~M.}\ \bibnamefont {Lalieu}}, \bibinfo {author} {\bibfnamefont {A.~J.~M.}\ \bibnamefont {Deenen}}, \bibinfo {author} {\bibfnamefont {R.~A.}\ \bibnamefont {Duine}},\ and\ \bibinfo {author} {\bibfnamefont {B.}~\bibnamefont {Koopmans}},\ }\href {https://doi.org/10.1103/PhysRevB.100.220409} {\bibfield  {journal} {\bibinfo  {journal} {Phys. Rev. B}\ }\textbf {\bibinfo {volume} {100}},\ \bibinfo {pages} {220409(R)} (\bibinfo {year} {2019})}\BibitemShut {NoStop}%
\bibitem [{\citenamefont {Iacocca}\ \emph {et~al.}(2019)\citenamefont {Iacocca}, \citenamefont {Liu}, \citenamefont {Reid}, \citenamefont {Z.Fu}, \citenamefont {Ruta}, \citenamefont {Granitzka}, \citenamefont {Jal}, \citenamefont {Bonetti}, \citenamefont {Gray}, \citenamefont {Graves}, \citenamefont {Kukreja}, \citenamefont {Chen}, \citenamefont {Higley}, \citenamefont {Chase}, \citenamefont {Guyader}, \citenamefont {Hirsch}, \citenamefont {Ohldag}, \citenamefont {Schlotter}, \citenamefont {Dakovski}, \citenamefont {Coslovich}, \citenamefont {Hoffmann}, \citenamefont {Carron}, \citenamefont {Tsukamoto}, \citenamefont {Kirilyuk}, \citenamefont {Kimel}, \citenamefont {Rasing}, \citenamefont {St{\"o}hr}, \citenamefont {Evans}, \citenamefont {Ostler}, \citenamefont {Chantrell}, \citenamefont {Hoefer}, \citenamefont {Silva},\ and\ \citenamefont {D{\"u}rr}}]{Iacocca2019}%
  \BibitemOpen
  \bibfield  {author} {\bibinfo {author} {\bibfnamefont {E.}~\bibnamefont {Iacocca}}, \bibinfo {author} {\bibfnamefont {T.-M.}\ \bibnamefont {Liu}}, \bibinfo {author} {\bibfnamefont {A.}~\bibnamefont {Reid}}, \bibinfo {author} {\bibnamefont {Z.Fu}}, \bibinfo {author} {\bibfnamefont {S.}~\bibnamefont {Ruta}}, \bibinfo {author} {\bibfnamefont {P.}~\bibnamefont {Granitzka}}, \bibinfo {author} {\bibfnamefont {E.}~\bibnamefont {Jal}}, \bibinfo {author} {\bibfnamefont {S.}~\bibnamefont {Bonetti}}, \bibinfo {author} {\bibfnamefont {A.}~\bibnamefont {Gray}}, \bibinfo {author} {\bibfnamefont {C.}~\bibnamefont {Graves}}, \bibinfo {author} {\bibfnamefont {R.}~\bibnamefont {Kukreja}}, \bibinfo {author} {\bibfnamefont {Z.}~\bibnamefont {Chen}}, \bibinfo {author} {\bibfnamefont {D.}~\bibnamefont {Higley}}, \bibinfo {author} {\bibfnamefont {T.}~\bibnamefont {Chase}}, \bibinfo {author} {\bibfnamefont {L.~L.}\ \bibnamefont {Guyader}}, \bibinfo {author} {\bibfnamefont {K.}~\bibnamefont {Hirsch}}, \bibinfo {author}
  {\bibfnamefont {H.}~\bibnamefont {Ohldag}}, \bibinfo {author} {\bibfnamefont {W.}~\bibnamefont {Schlotter}}, \bibinfo {author} {\bibfnamefont {G.}~\bibnamefont {Dakovski}}, \bibinfo {author} {\bibfnamefont {G.}~\bibnamefont {Coslovich}}, \bibinfo {author} {\bibfnamefont {M.}~\bibnamefont {Hoffmann}}, \bibinfo {author} {\bibfnamefont {S.}~\bibnamefont {Carron}}, \bibinfo {author} {\bibfnamefont {A.}~\bibnamefont {Tsukamoto}}, \bibinfo {author} {\bibfnamefont {A.}~\bibnamefont {Kirilyuk}}, \bibinfo {author} {\bibfnamefont {A.}~\bibnamefont {Kimel}}, \bibinfo {author} {\bibfnamefont {T.}~\bibnamefont {Rasing}}, \bibinfo {author} {\bibfnamefont {J.}~\bibnamefont {St{\"o}hr}}, \bibinfo {author} {\bibfnamefont {R.}~\bibnamefont {Evans}}, \bibinfo {author} {\bibfnamefont {T.}~\bibnamefont {Ostler}}, \bibinfo {author} {\bibfnamefont {R.}~\bibnamefont {Chantrell}}, \bibinfo {author} {\bibfnamefont {M.}~\bibnamefont {Hoefer}}, \bibinfo {author} {\bibfnamefont {T.}~\bibnamefont {Silva}},\ and\ \bibinfo {author}
  {\bibfnamefont {H.}~\bibnamefont {D{\"u}rr}},\ }\href {https://doi.org/10.1038/s41467-019-09577} {\bibfield  {journal} {\bibinfo  {journal} {Nat. Commun.}\ }\textbf {\bibinfo {volume} {10}},\ \bibinfo {pages} {1756} (\bibinfo {year} {2019})}\BibitemShut {NoStop}%
\end{thebibliography}%

\end{document}


\title{Magnon-mediated terahertz spin transport in metallic Gd$\vert$Pt stacks\\\,\\Supplemental Material}

\author{Oliver Gueckstock}
\email{Corresponding author: oliver.gueckstock@fu-berlin.de}
\affiliation{Fachbereich Physik, Freie Universit{\"a}t Berlin, Arnimallee 14,
	14195 Berlin, Germany}
\affiliation{Department of Physical Chemistry, Fritz Haber Institute of the Max Planck Society,  Faradayweg 4-6, 14195 Berlin, Germany}
\author{Tim Amrhein}
\affiliation{Fachbereich Physik, Freie Universit{\"a}t Berlin, Arnimallee 14,
14195 Berlin, Germany}
\author{Beatrice Andres}
\affiliation{Fachbereich Physik, Freie Universit{\"a}t Berlin, Arnimallee 14,
14195 Berlin, Germany}
\author{Pilar Jiménez-Cavero}
\affiliation{Centro Universitario de la Defensa, Academia General Militar, 50090 Zaragoza, Spain}
\author{Cornelius Gahl}
\affiliation{Fachbereich Physik, Freie Universit{\"a}t Berlin, Arnimallee 14,
	14195 Berlin, Germany}
\author{Tom S. Seifert}
\affiliation{Fachbereich Physik, Freie Universit{\"a}t Berlin, Arnimallee 14,
	14195 Berlin, Germany}
\author{Reza Rouzegar}
\affiliation{Fachbereich Physik, Freie Universit{\"a}t Berlin, Arnimallee 14,
	14195 Berlin, Germany}
\affiliation{Fachbereich Physik, Freie Universit{\"a}t Berlin, Arnimallee 14,
14195 Berlin, Germany}
\author{Ilie Radu}
\affiliation{Fachbereich Physik, Freie Universit{\"a}t Berlin, Arnimallee 14,
14195 Berlin, Germany}
\affiliation{European XFEL, Holzkoppel 4, 22869 Schenefeld, Germany}
\author{Irene Lucas}
\affiliation{Instituto de Nanociencia y Materiales de Aragón (INMA), Universidad de Zaragoza-CSIC, Mariano Esquillor, Edificio I+D, 50018 Zaragoza, Spain}
\affiliation{Departamento Física de la Materia Condensada, Universidad de Zaragoza, Pedro Cerbuna 12, 50009 Zaragoza, Spain}
\author{Marko Wietstruk}
\affiliation{Fachbereich Physik, Freie Universit{\"a}t Berlin, Arnimallee 14,
14195 Berlin, Germany}
\author{Luis Morellón}
\affiliation{Instituto de Nanociencia y Materiales de Aragón (INMA), Universidad de Zaragoza-CSIC, Mariano Esquillor, Edificio I+D, 50018 Zaragoza, Spain}
\affiliation{Departamento Física de la Materia Condensada, Universidad de Zaragoza, Pedro Cerbuna 12, 50009 Zaragoza, Spain}
\author{Martin Weinelt}
\affiliation{Fachbereich Physik, Freie Universit{\"a}t Berlin, Arnimallee 14,
14195 Berlin, Germany}
\author{Tobias Kampfrath}
\affiliation{Fachbereich Physik, Freie Universit{\"a}t Berlin, Arnimallee 14,
14195 Berlin, Germany}
\affiliation{Department of Physical Chemistry, Fritz Haber Institute of the Max Planck Society, Faradayweg 4-6, 14195 Berlin, Germany}
\author{Nele Thielemann-Kühn} 
\affiliation{Fachbereich Physik, Freie Universit{\"a}t Berlin, Arnimallee 14,
14195 Berlin, Germany}

\date{\today}

\pacs{Valid PACS appear here}
\keywords{Suggested keywords}
\maketitle
\section{Note 1: Sample preparation details}

\noindent
\textbf{1.1 Gd$\vert$Pt bilayers for THz emission experiments}\\
\noindent
The SiO$_2$\,(500\,µm)$\vert\vert$Y\,(5\,nm)$\vert$Gd\,(10\,nm)$\vert$Pt\,(2\,nm) samples were grown in the DynaMaX sample preparation cluster located at the PM3 beamline at HZB, BESSY II in Berlin. This cluster features a molecular beam epitaxy (MBE) chamber specialized on the growth of rare earth metals as well as a sputter deposition chamber for various other materials. The substrate, double-side polished polycrystalline SiO$_2$ with dimensions of (10\,x\,10\,x\,0.5)\,mm$^2$, was cleaned in an ultrasonic bath first in acetone and thereafter in ethanol for 15 minutes each. After transfer into the MBE chamber with a base pressure of $10^{-10}$\,mbar, the substrate was heated to 100 $^{\circ{}}$C for 10 minutes via indirect heating from a filament. To protect the Gadolinium (Gd) layer from oxidation by the substrate and to reduce strain imposed onto the Gd layer due to a lattice mismatch of SiO$_2$ and Gd, first, a 5\,nm Yttrium (Y) buffer layer was grown on top of the substrate with a growth rate of about $1\,–\,1.5\,$\AA/min. Subsequently, a 10 nm-thick Gd layer was grown on top with a rate of $1\,–\,2\,$\AA/min. For growth of the 2\,nm-thick Platinum (Pt) layer, the sample was transferred into the sputter deposition chamber at a base pressure of $10^{-9}$\,mbar. The Pt growth-rate was 0.1\,–\,0.2\,\AA/s at an Argon pressure of $8\cdot10^{-4}$-$1\cdot10^{-3}$\,mbar.
\newline
\newline
\noindent
\textbf{1.2 Sample preparation for photoelectron spectroscopy on Gd(0001)}\\
To measure the spin polarization of the photoelectrons emitted from Gd(0001), we used a spin detector based on exchange scattering  \cite{Winkelmann2008,Andres2015}. The electron scattering-target consists of an in-vacuo prepared 6\,monolayer Fe film on a W(001) single crystal substrate, passivated with a $p(1 \times 1)$ oxygen superstructure. Switching the in-plane magnetization of the Fe film between $\pm \bm{M}$ allows us to determine the spin polarization without changing the sample magnetization. We determined the Sherman function of the spin detector to 0.24 and obtained a figure of merit of $1.6 \cdot 10^{-3}$ at a scattering energy of 5.5\,eV.

\noindent
The  Gd(0001) sample was grown by molecular beam epitaxy, employing home-build evaporators.
For preparation we evaporated 100\,{\AA} of Gd onto a single crystal W(110) substrate at a temperature of 300\,K with a deposition rate of 5\,\AA/min.  The substrate was cleaned as described in Ref.~\cite{Andres_2015_2}. The pressure during evaporation was $3\cdot10^{-10}$\,mbar and the Gd film was annealed for 1\,min at 780\,K in order to obtain a well-ordered surface. Ferromagnetism and surface order are confirmed by the spin polarization of the Gd surface state and by low-energy electron diffraction, respectively \cite{Andres2015,Bobowski2024}.
\newline
\newline
\noindent
\textbf{1.3 $\gamma$-Fe$_2$O$_3$ $\vert$Pt for THz emission experiments}\\
\noindent
The $\gamma$-Fe$_2$O$_3$\,(10\,nm)$\mid$Pt\,(2.5\,nm) sample is epitaxially grown by pulsed laser deposition on a 500\,$\mu$m-thick MgO(001) substrate. A Fe$_3$O$_4$ target is used to deposit the maghemite layer in a two-step process. After deposition of the Fe$_3$O$_4$ layer, the sample was in-situ annealed at 325\,°C in an oxygen atmosphere that changed the sample’s phase to $\gamma$-Fe$_2$O$_3$. Subsequently, the maghemite film is in situ covered by DC sputtering with a 2.5\,nm-thick layer of Pt. More details on the growth process can be found in Ref. \cite{JimenezCavero2017}.

\newpage

\section{Note 2: Sample characterization}

\begin{figure}[h]
\includegraphics[width=\columnwidth]{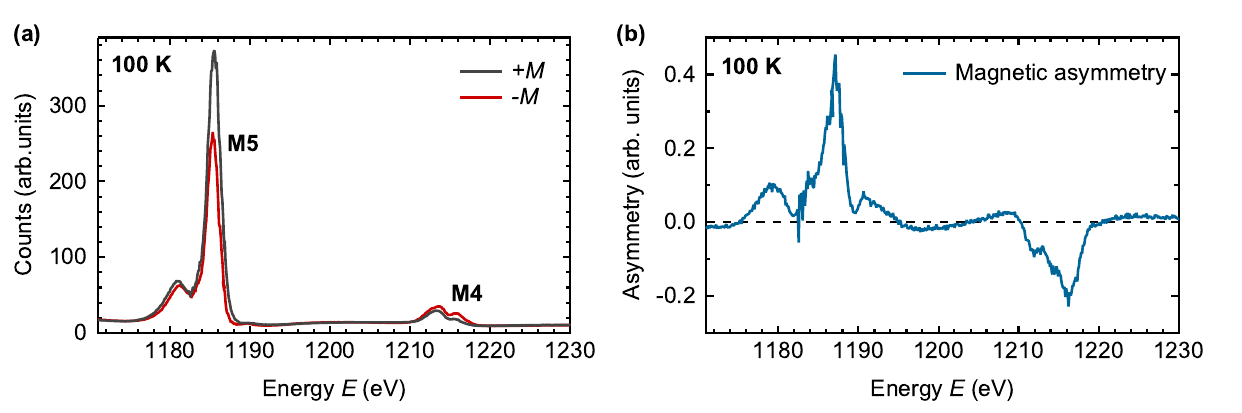}
\caption{X-ray magnetic circular dichroism of Gd. (a) X-ray scattering signals $S_{\rm{XMCD}}(E,M)$ of Gd for $M_5$ and $M_4$ absorption edges measured at a sample temperature of 100 K for oppositely oriented magnetic in-plane fields ($\pm \bm{M}$). (b) Magnetic asymmetry $S_{\rm{asym}}(E)$ from the XMCD measurement in panel (a) at the $M_5$ edge.}
\label{Figure_S1}
\end{figure}

\noindent
\textbf{2.1 X-ray magnetic circular dichroism - XMCD}\\
\noindent
To characterize the magnetic properties of the Gd$\mid$Pt sample, we performed a $(\theta,2\theta)$ X-ray reflection experiment at the PM3 beamline of BESSY II \cite{Kachel2015}. The incident X-ray beam was circularly polarized and hit the sample at a grazing angle of $5^\circ$ while the scattered light $S_{\rm{XMCD}}(E,M)$ was detected at $10^\circ$ with respect to the incident beam. A magnetic field $\bm{M}$ was applied along the intersection of the sample and the scattering plane. Spectra displayed in Fig.\,\ref{Figure_S1}(a) were recorded at the Gd $M_5$ and $M_4$ edges for opposite orientations of the magnetic field $\pm \bm{M}$ at a sample temperature of 100 K. A clear magnetic contrast can be seen between the two spectra at both absorption edges. The magnetic asymmetry was calculated as
\begin{equation}
    S_{\rm{asym}} (E) = S_{\rm{XMCD}}(E,+M)-S_{\rm{XMCD}}(E,-M)/S_{\rm{XMCD}}(E,+M)+S_{\rm{XMCD}}(E,-M)
\end{equation}
and is shown in Fig.\,\ref{Figure_S1}(b). At the Gd $M_5$ edge, the magnetic asymmetry increases to about 45\%, which proves that the Gd thin film is highly magnetic.

\newpage

\begin{figure}[h]
\includegraphics[width=0.45\textwidth]{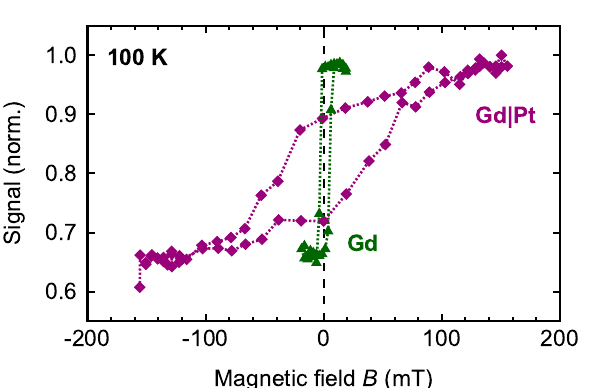}
\caption{Magnetic hysteresis loops of Gd (green) and Gd$\vert$Pt (violet) recorded at 100\,K sample temperature. The Gd$\vert$Pt thin-film sample on a glass substrate exhibits a significantly larger coercive field than pure Gd.}
\label{Figure_S2}
\end{figure}

\noindent
\textbf{2.2 Magnetic hysteresis}\\
\noindent
The magnetic hysteresis loop for the Gd$\vert$Pt sample was recorded at the Gd $M_5$ edge. As displayed in Fig.\,\ref{Figure_S2} (violet squares) the coercive field amounts to about 60\,mT. For comparison, the green triangles in Fig.\,\ref{Figure_S2} show a hysteresis loop recorded for a 10-nm thick Gd film on a W(110) substrate. The coercive field of the pure Gd film is with 10\,mT much lower than that of Gd$\vert$Pt. This indicates that the Gd magnetization in the Gd$\vert$Pt sample is stabilized by exchange interaction at the Gd$\vert$Pt interface.

\newpage
\section{Note 3: Terahertz emission experiment}

\noindent
\textbf{3.1 Temperature dependence of the THz signal from Gd$\mid$Pt}

\begin{figure}[h]
\includegraphics[width=0.55\columnwidth]{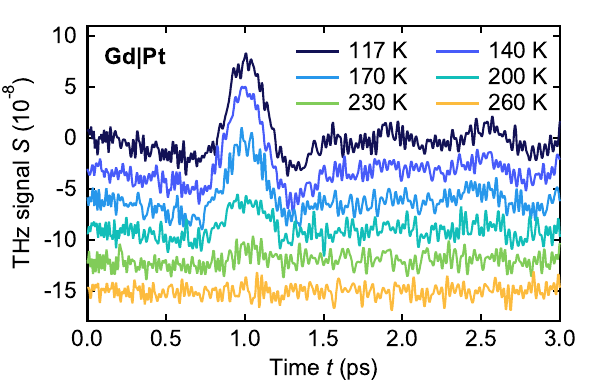}
\caption{Temperature-dependent raw data of Gd$\mid$Pt bilayers. Electro-optic signal $S(t)$ from optically excited Gd$\mid$Pt as function of the sample temperature $T_0+\Delta T_0$ ranging from 117\,K to 260\,K. Here, $T_0$ is the set temperature, and $\Delta T_0$ is the estimated stationary temperature increase due to the continuous sample heating by the pump beam as detailed in Note 3.2.}
\label{Figure_S3}
\end{figure}

\noindent
\textbf{3.2 Estimation of stationary temperature increase}\\
\noindent
For Permalloy (Py)-based bilayer systems on glass substrates with typical total thickness of 5\,nm we estimate the stationary temperature increase due to the high repetition rate of laser pulses to $\Delta T_0\sim80$-$120$\,K. This estimate is based on previous experiments varying the pump fluence of Py$\vert$Pt bilayers. Here, in case of Gd$\vert$ Pt, we need to consider the higher thickness of Gd (10\,nm) and its lower electronic heat capacity \cite{Griffel1954,Yokokawa1979}. By comparing these values to a reference sample of Py\,(3\,nm)$\mid$Pt\,(3\,nm), we, thus, estimate a lower stationary temperature increase of $\Delta T_0\sim35$-$45$\,K for Gd$\vert$Pt on a glass substrate. We note that $\Delta T_0$ also depends on temperature since also the heat capacities are temperature-dependent. Due to a high uncertainty of $\Delta T_0$, we take the mean value of $40$\,K to correct all temperatures $T_0$.  This estimation is a lower bound of $\Delta T_0$.\\

\newpage

\noindent
\textbf{3.3 Temperature dependence of spin current dynamics from Gd$\mid$Pt}

\begin{figure}[h]
\includegraphics[width=.55\columnwidth]{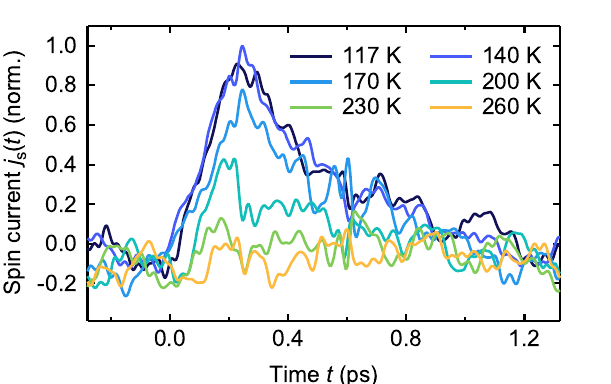}
\caption{Extracted spin current dynamics of Gd$\mid$Pt bilayers for various sample temperatures $T_0+\Delta T_0$ ranging from 117\,K to 260\,K. Up to the point, where the signal-to-noise ratio permits a reasonable judgement of the dynamics, $j_s(t,T)$ is not significantly changing. }
\label{Figure_S4}
\end{figure}
\noindent
\textbf{3.4 Estimation of the temperature gradient of electrons in Gd$\mid$Pt}\\
To estimate the temperature gradient of electrons between metallic Gd with a thickness of 10\,nm and 2\,nm-thick Pt, we calculate the temperature change $\Delta T_i$ of the individual layers by
\begin{equation}
    \Delta Q_i=\Delta T_i m c_e
\end{equation}
with  $\Delta Q_i$ being the amount of heat, $m$ and $c_e$ denoting the mass and electronic heat capacity, respectively. We assume that the electric field of the pump pulse is constant across the entire sample stack and therefore, also in every individual layer. Then, $\Delta Q_i$ is proportional to the absorbance $A$, which can be considered to be proportional to $\textrm{Im}(n^2)$ with $n$ the refractive index. The mass of each layer can be described by $m\propto d\cdot\rho$, where $d$ is the layer thickness and $\rho$ the material density. In total, the temperature change upon optical excitation can be described by
\begin{equation}
    \Delta T_i\propto \frac{\textrm{Im}(n_i^2)}{d_i \rho_i c_{e,i}}
\end{equation}
with $i$=Gd or Pt. By normalizing to $\Delta T_{\textrm{Gd}}$ we estimate the electronic temperature increase inside the Gd layer to be half of that in Pt ($\Delta T_{\textrm{Gd}}\approx0.5\Delta T_{\textrm{Pt}}$). Used material parameters can be found in Table \ref{tab:S1}.\\\\
\noindent
We further note that the thermalization of Pt electrons upon optical excitation is not severely altered by the presence of a metallic Gd layer because the electron-phonon equilibration rate $\Gamma_\text{ep}=g_\text{ep}/c_\text{e}$ \cite{Rouzegar2022} is very similar for Gd ($3.75\cdot 10^{12}$) and Pt ($2.75\cdot 10^{12}$). Here $g_\text{ep}$ and $c_\text{e}$ are the electron-phonon coupling parameter and electronic heat capacity, respectively.  Values are taken from Refs. \cite{Frietsch2016,Bobowski2017}.\\
\begin{table}[h]
    \centering
    \begin{tabular}{ccccccc}
        \hline
        \textbf{Material}&$\boldsymbol{n}$&\textbf{Im($\boldsymbol{n}^2$)}&\textbf{$\boldsymbol{c_e}$ [J/(kg K)]}&\textbf{$\boldsymbol{d}$ [nm]}&\textbf{$\boldsymbol{\rho}$ [g/cm$^3$]}&\textbf{$\boldsymbol{\Delta T}$ [norm.]}\\
        \hline 
        Gd&$2.55+2.99\textrm{i}\cite{Akashev2019}$&15.25&163.7\cite{Griffel1954}&10&7.901&0.48\\
        Pt&$0.58+8.08\textrm{i}$ \cite{Werner2009}&9.31&87,6 \cite{Yokokawa1979}&2&21.45&1\\
        Fe&/&/&364.9 \cite{Desai1986}&typ. 3 nm&/&/ 
    \end{tabular}
    \caption{Material parameters of Gd, Pt and Fe. The specific electronic heat capacities are given for temperatures at 75\,K for Gd, 80\,K for Pt and 70\,K for Fe. The temperature change $\Delta T$ is normalized to the value of Pt.}
    \label{tab:S1}
\end{table}

\newpage

\section{Note 4: Pump-probe photoemission experiment}

\begin{figure}[h]
\includegraphics[width=0.55\columnwidth]{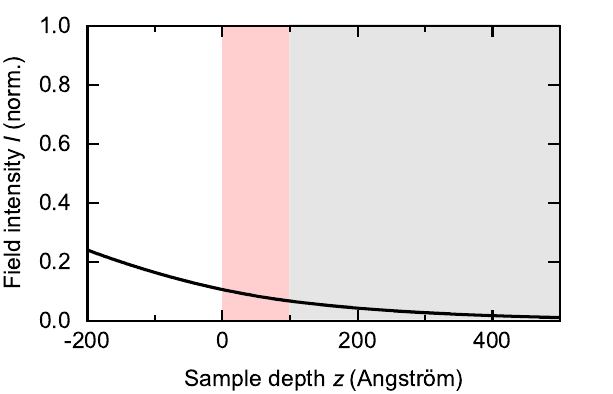}
\caption{Absorbed field intensity $I$ in Gd/W(110) as a function of sample depth $z$.}
\label{Figure_GdPt_FigS5_v2025-02-11}
\end{figure}

\noindent
{\bf 4.1 Absorbed pump fluence}\\
\noindent
The Gd/W(110) sample was excited by $s$-polarized 800\,nm laser pulses at an incidence angle of $45^{\circ}$. We calculated the absorbed laser intensity using the program IMD by D.~Windt \cite{IMD_1998} in conjunction with the XOP (X-ray oriented programs) driver software \cite{XOP_1997}. Figure~\ref{Figure_GdPt_FigS5_v2025-02-11} shows the electric field intensity $I$ of the pulse distributed over the sample depth $z$. The red marked area depicts the Gd layer, while the gray area stands for the W substrate. The field intensity is modulated by interference between the incoming and reflected beam at each interface and forms a standing wave pattern in front of the sample. As s-polarized light oscillates parallel to the sample interfaces and the divergence of the electric field part of the light must be zero, the field intensity is constant when crossing an interface. 62\% of the incident intensity is reflected at the Gd metal surface. We determined an absorption of only 4\% of the incident intensity in the 10-nm Gd layer, while the remaining 34\% are absorbed in the W substrate.\\
\noindent
At the pump fluence of 8\,mJ/cm$^2$, the average absorbed energy density of the 10-nm Gd film corresponds to 330\,kJ/cm$^3$.

\begin{figure}[h]
\includegraphics[width=0.5\columnwidth]{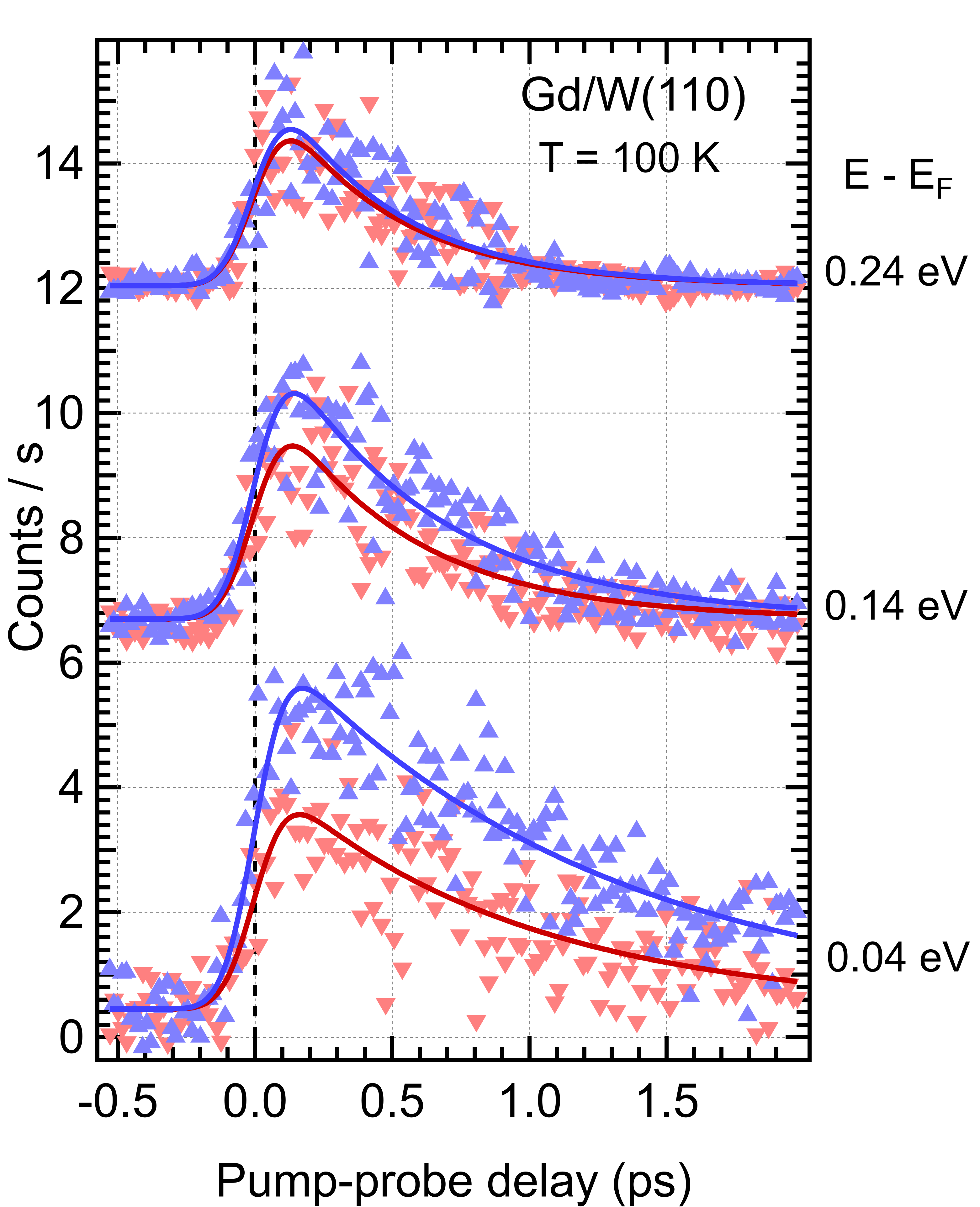}
\caption{Time-resolved photoemission traces for spin majority and spin minority electrons from epitaxial Gd(0001) on W(110) at ener-
gies $E-E_{\rm{F}}$ above the Fermi level as indicated.}
\label{Figure_S5}
\end{figure}

\begin{figure}[h]
\includegraphics[width=0.5\columnwidth]{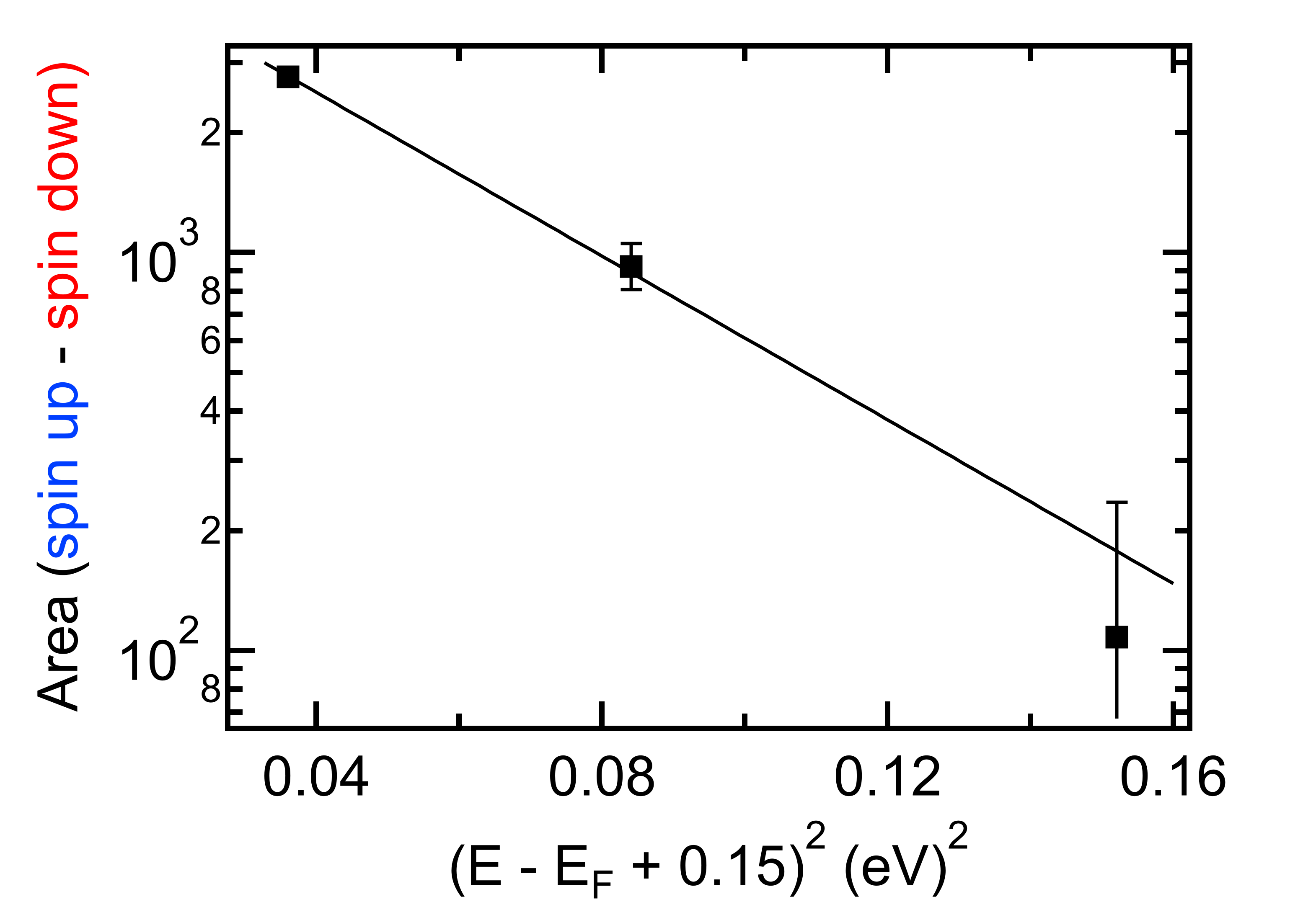}
\caption{Difference in total count rate of spin-up and spin-down channels of Fig.\,\ref{Figure_S5} on a semi-logarithmic scale as a function of the square of the energy above the surface state peak-maximum $(\Delta E)^2 = (E-E_{\rm{F}} + 0.15)^2$.}     
\label{Figure_S6}
\end{figure}

\noindent
\textbf{4.2 Time-resolved photoemission at energies close to the Fermi level}\\
Figure\,\ref{Figure_S5} shows photoemission traces vs pump-probe delay for spin majority (blue, up-pointing triangle) and spin minority (red, down-pointing triangle) electrons at energies $E-E_{\rm{F}}$ above the Fermi level. The traces close above the Fermi level recorded at 0.04\,eV, 0.14\,eV and 0.24\,eV show a spin polarization of majority character, which decreases with increasing distance from the Fermi level. At energies ($E-E_{\rm{F}} \geq 0.3$\,eV), there is no significant spin polarization observable (see Fig. 4, main text). 
The finite positive spin polarization at low energies therefore points to an additional source, the occupied surface state below $E_{\rm{F}}$ \cite{Bode1999,Andres2015}. It is well described by a Gaussian centered at $E_{\rm{F}} - E = 0.15$\,eV binding energy.
If we assume that the spin polarization in Fig.\,\ref{Figure_S5} originates from this surface state, the difference in count rates of spin up and spin down channels must follow its Gaussian shape $\propto \exp{\Delta E^2}$ \cite{Andres2015}.
Figure\,\ref{Figure_S6} shows on a semi-logarithmic plot this difference in spin-dependent count rates as a function of $(\Delta E)^2 = (E-E_{\rm{F}} + 0.15)^2$, \textit{i.e.}, the square of the energy with respect to the peak maximum of the surface state. The dependence indicates that the bulk component contributes negligible spin polarization even at low energies.

\newpage

\bibliographystyle{apsrev4-2}
\bibliography{GdPt}